\documentclass{aa}  

\usepackage{natbib} 
\usepackage{graphicx}
\usepackage{hyperref}
\hypersetup{colorlinks=true,linkcolor=blue,citecolor=blue,urlcolor=blue}

\usepackage{txfonts}
\usepackage{siunitx}
\usepackage{physics}
\usepackage{booktabs}
\usepackage{comment}
\def\alstar{\textsc{AlStar}}
\def\starlight{\textsc{starlight}}
\newcommand{\hii}{H\thinspace\textsc{ii}}
\newcommand{\Ha}{\ifmmode \mathrm{H}\alpha \else H$\alpha$\fi}
\newcommand{\Hb}{\ifmmode \mathrm{H}\beta \else H$\beta$\fi}
\newcommand{\nii}{\ifmmode [\mathrm{N}\,\textsc{ii}] \else [N~{\scshape ii}]\fi}
\newcommand{\oiii}{\ifmmode [\mathrm{O}\,\textsc{iii}] \else [O\,{\scshape iii}]\fi}
\newcommand{\sii}{\ifmmode [\mathrm{S}\,\textsc{ii}] \else [S~{\scshape ii}]\fi}
\newcommand{\oi}{\ifmmode [\mathrm{O}\,\textsc{i}] \else [O\,{\scshape i}]\fi}
\newcommand{\oii}{\ifmmode [\mathrm{O}\,\textsc{ii}] \else [O\,{\scshape ii}]\fi}
\newcommand{\nOne}{\ifmmode [\mathrm{N}\,\textsc{i}] \else [N~{\scshape i}]\fi}
\newcommand{\siii}{\ifmmode [\mathrm{S}\,\textsc{iii}] \else [S~{\scshape iii}]\fi}

\newcommand{\HaNii}{\ifmmode \mathrm{H}\alpha\mathrm{N}\textsc{ii} \else H$\alpha$N{\scshape ii}\fi}
\newcommand{\WHaNii}{\ifmmode \mathrm{W}_{\mathrm{H}\alpha\mathrm{N}\textsc{ii}} \else W$_{\mathrm{H}\alpha\mathrm{N}\textsc{ii}}$\fi}
\newcommand{\WHaN}{\ifmmode W_{{\mathrm H}\alpha{\rm N}} \else $W_{{\mathrm H}\alpha{\mathrm N}}$}
\newcommand{\WHa}{\ifmmode W_{{\mathrm H}\alpha} \else $W_{{\rm H}\alpha}$}

\usepackage{xcolor}
\definecolor{Jpurple}{RGB}{ 155, 0, 255 }

\begin{document} 
   \title{Spatially resolved stellar populations and emission lines properties in nearby galaxies with J-PLUS}

   \subtitle{I. Method and first results for the M101 group}

   \author{J. Thainá-Batista\inst{1,2}, R. Cid Fernandes\inst{2}, R. M. González Delgado\inst{1}, J.E. Rodríguez-Martín\inst{1}, R. García-Benito\inst{1}, G. Martínez-Solaeche\inst{1}, L. A. Díaz-García\inst{1}, V. H. Sasse\inst{2}, A. Lumbreras-Calle\inst{3}, A. M. Conrado\inst{1},  
   J. Alcaniz\inst{5}, R. E. Angulo\inst{8,9}, A. J. Cenarro\inst{3,4}, D. Cristóbal-Hornillos\inst{3}, R. A. Dupke\inst{5}, A. Ederoclite\inst{3,4}, C. Hernández-Monteagudo\inst{6,7}, C. López-Sanjuan\inst{3,4}, A. Marín-Franch\inst{3,4}, M. Moles\inst{3}, L. Sodré Jr.\inst{10}, H. Vázquez Ramió\inst{3,4}, \and J. Varela\inst{3}
          }

    \titlerunning{The M101 group with J-PLUS}
    \authorrunning{Thainá-Batista \& the J-PLUS collaboration}

   \institute{Instituto de Astrofísica de Andalucía (CSIC), PO Box 3004, 18080 Granada, Spain\\
              \email{jullia.thainna@gmail.com}
         \and
             Departamento de Física, Universidade Federal de Santa Catarina, PO Box 476, 88040-900 Florianópolis, SC, Brazil
        \and 
        Centro de Estudios de Física del Cosmos de Aragón, Plaza San Juan 1, 44001 Teruel, Spain
        \and
            Unidad Asociada CEFCA-IAA, CEFCA, Unidad Asociada al CSIC por el IAA, Plaza San Juan 1, 44001 Teruel, Spain
        \and
            Observatório Nacional - MCTI (ON), Rua Gal. José Cristino 77, São Cristóvão, 20921-400 Rio de Janeiro, Brazil
        \and 
            Instituto de Astrofísica de Canarias, La Laguna, 38205, Tenerife, Spain
        \and Departamento de Astrofísica, Universidad de La Laguna, 38206, Tenerife, Spain
        \and Donostia International Physics Centre (DIPC), Paseo Manuel de Lardizabal 4, 20018 Donostia-San Sebastián, Spain
        \and IKERBASQUE, Basque Foundation for Science, 48013, Bilbao, Spain
        \and Instituto de Astronomia, Geofísica e Ciências Atmosféricas, Universidade de São Paulo, São Paulo, SP 05508-090, Brazil
             }


  \abstract
   {Spatially resolved maps of stellar populations and nebular emission are key tools for understanding the physical properties and evolutionary stages of galaxies. These maps are commonly derived from Integral Field Spectroscopy (IFS) data or, alternatively, from multi-band imaging techniques.
   }
   {
   We aim to characterize the spatially resolved stellar population and emission line properties of galaxies in the M101 group using Javalambre Photometric Local Universe Survey (J-PLUS) data.
   }
   {
   The datacubes first go through pre-processing steps which include masking, noise suppression, PSF homogenization, and spatial binning. The improved data are then analyzed with the spectral synthesis code \alstar, which has been previously shown to produce excellent results with the unique 12 bands filter system of J-PLUS and S-PLUS.
   }
   {
   We produce maps of stellar mass surface density ($\Sigma_\star$), mean stellar age and metallicity, 
   star formation rate surface density ($\Sigma_{\rm SFR}$), dust attenuation, and emission line properties such as fluxes and equivalent widths of the main optical lines. Relations among these properties are explored.
    All galaxies exhibit a well defined age-$\Sigma_\star$ relation, except for the dwarfs.
    Similarly, 
    all of the galaxies follow local $\Sigma_\star$-$\Sigma_{\rm SFR}$ star-forming MS relations, with specific star formation rates that grow for less massive systems. 
    A stellar $\Sigma_\star$-metallicity relation is clearly present in M101, while other galaxies have either flatter or undefined relations. Nebular metallicities correlate with $\Sigma_\star$ for all galaxies.
   }
   {
   This study demonstrates the ability of J-PLUS to perform IFS-like analysis of galaxies, offering robust spatially resolved measurements of stellar populations and emission lines over large fields of view. The M101 group analysis showcases the potential for expanding such studies to other groups and clusters, contributing to the understanding of galaxy evolution across different environments.
   }
   \maketitle

\section{Introduction}
\label{sec:Intro}
Galaxy groups are the smallest aggregates of galaxies, typically composed of fewer than a hundred galaxies, representing the smallest structures that collapse to form galaxy clusters \citep{PressSchechter1974}. 
Groups are common in the nearby Universe, containing approximately half of the galaxies in this volume \citep{Eke2004}. They represent an intermediate environment between galaxies in the field and those in dense clusters. 

Groups are associated with a range of galaxy properties that reflect the diversity of processes acting at these scales \citep{Wetzel2012}. 
Comparisons between galaxies in groups, clusters, and the field have revealed systematic differences in star formation rates (SFRs), gas content, metallicity, and structural properties (e.g. \citealt{Blanton2009}).

Recently, miniJPAS \citep{2021Bonoli} has illustrated the power of the J-PAS survey \citep{Benitez2014_JPAS} to detect groups with masses of up to $10^{13}$ M$_\odot$ and to characterize their galaxy populations up to a redshift of $z \sim 1$ \citep{Maturi2023, Doubrawa2024}. Its multiwavelength photometry system, with 56 narrow-band filters covering the whole optical range, allows the identification and characterization of galaxy populations and their evolution since $z = 1$ \citep{GonzalezDelgado2021, DiazGarcia2024}, as well as the stellar population properties of group members \citep{RodriguezMartin2022, GonzalezDelgado2022}. 

Similarly, the PAU \citep{Csizi2024PAU} and ALHAMBRA \citep{DiazGarcia2019, DiazGarcia2015MUFFIT} surveys have also demonstrated the capability of consecutive narrow-band filters to determine the stellar population properties of galaxies.  Naturally, one does not reach the accuracy obtainable with spectroscopic data (see, for example, \citealt{Nersesian2024, Nersesian2025}), but the results are useful nonetheless.

These recent studies, as well as most previous ones \citep[e.g.,][]{Boselli2006}, 
used integrated light to examine galaxy properties. 
Judging from the progress achieved when integrated-light surveys like the SDSS \citep{York2000SDSS} were complemented with Integral Field Spectroscopy (IFS) surveys like CALIFA \citep{Sanchez2012CALIFA} and MaNGA \citep{Bundy2015MaNGA}, examining the spatially resolved stellar and nebular properties of galaxies in groups is a worthwhile endeavor. Observationally, the challenge here is to cover galaxies over their full spatial extent as well as wide areas of the sky to map all group members. Additionally, statistics over many groups is needed to firmly establish whether galaxies in groups in different stages of evolution differ in their internal stellar and nebular properties.
These requirements are not easy to satisfy with IFS.

This paper is the first in a series aimed at characterizing the spatially resolved stellar population and emission line (EL) properties of galaxies in groups and clusters with J-PLUS \citep{Cenarro2019jplus} and S-PLUS \citep{2019Splus, Herpich2024} data with the methodology developed by \cite{ThainaBatista2023} (hereinafter TB23). 
The large field of view of these surveys makes them ideally suited to study groups and clusters, while their unique combination of narrow and broad band filters offers information on both stellar and nebular properties.

This work focuses primarily on describing the data, the methodology, and an initial analysis of the results for the six galaxies in the M101 group \citep{Garcia1993}. Subsequent papers will target other nearby groups and clusters, such as the M51 group and Fornax \citep{SmithCastelli2024Fornax}, gradually building a sample suitable for comparative analyses of galaxy properties across different environments.
Sec.\ \ref{sec:sample} describes the sample and the pre-processing procedures applied to the data. Sec.\ \ref{sec:data_analysis} reviews and illustrates the fitting technique used. Our primary 2D results are presented in Sec.\ \ref{sec:Results_M101group}, where we examine maps of stellar population and EL properties, as well as relations between them. Sec.\ \ref{sec:Discussion} compares the galaxies in the group with one another in terms of their scaling relations. Finally, our main results are summarized in Sec.\ \ref{sec:conclusions}.

\section{Data, sample, and pre-processing}
\label{sec:sample}

This work explores the properties of galaxies in the M101 group, which, besides M101, harbors NGC 5474, NGC 5585, NGC 5204, UGC 8837, and NGC 5477. After a brief description of the data (Sec. \ref{sec:data}) and the individual galaxies in the sample (Sec. \ref{sec:IndividObjects}), we explain a series of pre-processing steps applied to the data  (Sec. \ref{sec:PreProcessing}) prior to the stellar population and EL analysis.

\subsection{Data}
\label{sec:data}

All six galaxies were observed as part of the Javalambre-Photometric Local Universe Survey (J-PLUS) with an 80 cm robotic telescope in the Observatorio Astrofísico de Javalambre \citep{oaj}, in Teruel, Spain. The Javalambre Auxiliary Survey Telescope (JAST80; \citealt{t80cam}) is equipped with a $9216 \times 9232$ pix camera (T80Cam) with a $1.4 \times 1.4\deg^2$  field of view and a pixel scale of 0.55 arcsec pix$^{-1}$. In this work, we use the third data release (DR3; \citealt{clsj24gphot}) of J-PLUS.
The distinctive feature of J-PLUS is its set of 12 filters, with five broad and seven narrow bands, covering the $\sim 3500$–9000 \AA\  range. The filters were designed for a multi-purpose survey, spanning applications ranging from the Solar system to extragalactic scales, see \cite{Cenarro2019jplus} for a full description of J-PLUS (including details on the observations and data reduction), and \cite{Logrono2019}, \cite{SanRoman2019}, \cite{2022_Lumbreras-Calle}, \cite{2022_Lopez-Sanjuan} and \cite{Rahna2025} for a few example studies based on its data.

\begin{figure*}
    \centering
    \includegraphics[width=\textwidth]{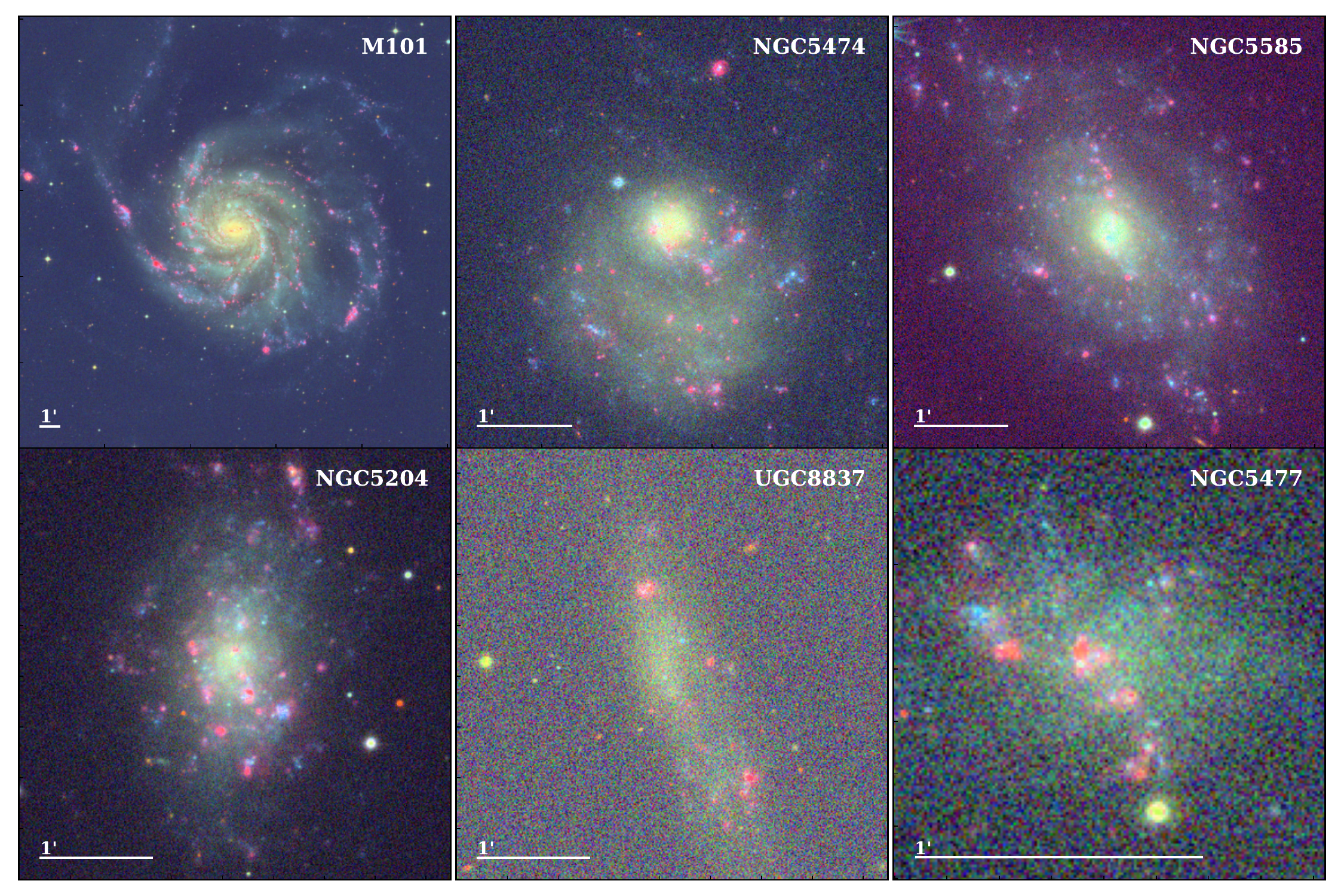}
    \caption{Composites of the original data for galaxies in the M101 group, built using the \textit{J0660}, g, and the sum of the five bluest filters in the R, G, and B channels, respectively. The white bars are 1 arcmin wide. }
    \label{fig:groupM101_original}
\end{figure*}

Fig.~\ref{fig:groupM101_original} shows J-PLUS images of all six galaxies in the group. These composite images are built using the \textit{J0660} flux in the R channel, the \textit{g} band in the G channel, and the sum of the fluxes in the five bluest bands (\textit{u, J0378, J0395, J0410, J0430}) in the B channel. As these images reveal, the group consists of spiral and irregular galaxies,  with sizes ranging from $\sim 1$ to 18 arcmin. The figure also reveals that all galaxies are asymmetric to some degree, and, as evidenced by the red regions in these images, they all exhibit \Ha\ emission, which falls well within the \textit{J0660} filter\footnote{Which is centered in 6600 \AA\ with a width of $\sim 138$ \AA.} for the redshifts of the galaxies in the group.

\subsection{Individual objects}
\label{sec:IndividObjects}

Table \ref{tab:sample} lists basic information on our sample galaxies. 
We adopt a single distance of 7.24 Mpc to all galaxies in the group, as determined by \cite{2012LeeJang} for M101. Information regarding morphologies was extracted from LEDA \citep{leda2014Makarov}, NED \citep{ned1990Helou}, and SIMBAD \citep{simbad2000Wenger}.

\begin{table}[h!]
\centering
\caption{Galaxies in our sample.}
\begin{tabular}{lccc}
\toprule[1.5pt]
Galaxy & Morphology & $\log M_\star/M_\odot$ & $M_r$ \\
\midrule[1.5pt]
M101     &  SAB(rs)cd.    &  10.35&  -21.23 \\
NGC 5474 &  SA(s)cd       &  9.05 &  -18.29 \\
NGC 5585 &  SAB(s)d       &  8.91 &  -18.12 \\
NGC 5204 &  SA(s)m        &  8.74 &  -17.89 \\
UGC 8837 &  IB(s)m        &  8.03 &  -15.97 \\
NGC 5477 &  SA(s)m or I   &  7.61 &  -15.38 \\
\bottomrule[1.5pt]
\end{tabular}
\tablefoot{The stellar mass and absolute AB r-band magnitude are derived from our own J-PLUS datacubes (Sec. \ref{sec:Results_M101group}).}
\label{tab:sample}
\end{table}

{\smallskip\noindent M101:}
Popularly known as the Pinwheel galaxy, M101 (also named NGC 5457, PGC 50063, UGC 08981) is the dominant galaxy of the group. With a redshift of 0.000804, this $\sim$ face-on, lopsided spiral of nearly 30 arcmin optical diameter exhibits tidal features as well as plenty of star forming regions, including giant \hii\ regions of nearly 1 kpc in size \citep{GarciaBenito2011}. Due to these characteristics, M101 is the subject of many studies in the literature --- see, e.g., \cite{2007Bresolin}, \cite{2018Hu}, \cite{Garner_2024} and references therein.

{\smallskip\noindent NGC 5474:} This dwarf spiral (also known as PGC 50216) also exhibits a lopsided structure, a feature often attributed to its gravitational interaction with M101 \citep{2022Liden}. NGC 5474 is the most massive satellite of M101, and has a redshift of 0.000874. Its peculiar structure and the presence of regions with ongoing star formation make it an interesting example of tidal effects within galaxy groups.

{\smallskip\noindent NGC 5585:}
Also referred to PGC 51210, this is a late-type galaxy of morphological type SAB(s)d with a redshift of 0.001011. It is particularly notable for being a dark-halo-dominated galaxy \citep{1991Cote}. 

{\smallskip\noindent NGC 5204:}
This low surface brightness galaxy (also known as PGC 47368) has a redshift of 0.000670. Like the other galaxies in the group, NGC 5204 exhibits an asymmetric structure, with numerous regions of star formation.

{\smallskip\noindent UGC 8837:} This irregular galaxy also referred to as PGC 49448, has a redshift of 0.00048. It is the only edge-on galaxy in the group. 

{\smallskip\noindent NGC 5477:}
With a redshift of 0.00105, this irregular dwarf galaxy (also known as PGC 50262) is the smallest in the M101 group, with a diameter of approximately 1 arcmin in the \textit{R}-band. Its spectrum is characterized by strong emission lines that indicate ongoing star formation. 

\subsection{Pre-processing}
\label{sec:PreProcessing}

The original flux-calibrated J-PLUS images underwent a series of pre-processing steps before being analyzed for their stellar population and EL information content. 
The general goals of these pre-processing steps are to clean the data of unwanted or spurious sources (such as foreground stars and background galaxies), reduce the noise level, homogenize the spatial resolution, and ensure photo-spectra with a minimal quality to warrant an analysis with the tools detailed in Sec. \ref{sec:data_analysis}. The main steps are summarized as follows:

\begin{enumerate} 
    \item {Spatial masks} - 
    The first step is to define spatial masks mapping the (a) object of interest, (b) foreground stars and background galaxies, and (c)  the sky. The ``stars'' mask (which includes foreground stars, background galaxies, and artifacts)
    was built using functions from the {\scshape photutils} package \citep{larry_bradley_2024_photutils} combined with our own code to detect sources, followed by an interactive review to confirm source identification and to add saturated stars. The galaxy limits were defined using isophotes at \textit{r}-band surface brightness $\mu_r = 24$\ mag/arcsec$^2$, allowing for 
    adjustments when necessary.

    \item {Butterworth filter} - Inspired by the work of \cite{Ricci_2014}, we apply a low-pass Butterworth filter that reduces noise and instrumental artifacts, particularly in the outer regions, where the signal is weaker. 

    \item {PSF homogenization} -  The full width at half maximum (FWHM) for the different bands is then used to convolve all images to the worst point source function (PSF), resulting in a common spatial resolution. 

    \item {Binning} - We then rebin the data on $2 \times 2$ pixels, changing the original 0.55 arcsec/pix scale to 1.1 arcsec/pix. This step increases the signal-to-noise (S/N) and optimizes computational operations while keeping the spatial resolution close to that of the original data.
    
    \item {Pixel tagging} - Next, we assign each pixel to one of three categories: (a) \hii\ region, (b) dusty region, and (c) unlabeled region. 
    Pixels tagged as \hii\ region are selected as those with a combined \Ha\ + \nii\ equivalent width of $> 40$ \AA\ and a flux $> 5.5\sigma$ above the average over the whole galaxy.
    The \Ha\ + \nii\ values used in this stage are obtained with the method of  \cite{2015_Vilella_Rojo} (see also \citealt{Rahna2025}).
    Dusty pixels are defined as those with a $z$-band/$g$-band flux ratio greater than 1.2, a condition which does not happen for any galaxy in this work.
    These tags are used in the next step (Voronoi binning), which is carried out separately for each tag, thus mitigating the mixing of physically different kinds of pixels within the same bin. We postpone a graphical illustration of this tagging strategy to the case of M51 (Thainá-Batista et.\ al., in prep.), which has plenty of both \hii\ and dusty regions.

    \item {Voronoi Binning} - Finally, we apply Voronoi binning \citep{2003Cappelari_Voronoi} to ensure a minimal S/N across bins. After testing different target S/N and filter combinations, we selected the \textit{u}-band as the reference band with a specific S/N for each galaxy: S/N$_u=8$ for NGC5477 and UGC 8837, S/N$_u=10$ for M101, NGC 5474 and NGC 5204, and S/N$_u=20$ for NGC 5585.

    \item {Local sky re-subtraction} - If the sky regions present a systematically net positive spectrum, then we find it worth refining the original sky subtraction by subtracting the average sky spectrum in the datacube. This optional extra step affects only the fainter regions of the galaxy.
    
\end{enumerate}

Fig.\ \ref{fig:pre-processing_steps_M101} illustrates the progressive effects of these steps, while Fig.\ \ref{fig:groupM101_posprocessing} shows how Fig. \ref{fig:groupM101_original} looks like after the pre-processing. 
The final datacubes are divided into 11706, 752, 1512, 2753, 120, and 133 Voronoi zones for M101, NGC5474, NGC5585, NGC5204, UGC8837, and NGC5477, respectively, covering areas corresponding to equivalent circular radii ranging from 28 kpc (M101) to 2 kpc (NGC 5477).

The errors in the photometric fluxes ($\epsilon_\lambda$) are important both in step 6 (Voronoi binning) and in the photo-spectral fits discussed below.  
A simple propagation of errors produces S/N ratios in excess of 100 in the \textit{r} band, a level of precision that one cannot realistically hope to achieve given the simplifications and limitations in the spectral modeling of galaxies, including our own.
We have thus limited the errors in the \textit{r} band such that S/N$_r$ does not exceed a maximum of 25. 
Errors in the other bands are adjusted in order to preserve the shape of the error spectrum of the original data, as computed from the median $\epsilon_\lambda / \epsilon_r$ spectrum within the galaxy mask. This spectrum is computed for each galaxy separately, though it is very similar in all cases, with errors in the five bluest bands on average $7.7 \times \epsilon_r$, while those in the four reddest bands are $1.6 \times \epsilon_r$.

\section{Data Analysis}
\label{sec:data_analysis}

The pre-processed photo-spectra are fitted using the \textsc{AlStar} code, a SED fitting tool similar to others like  \textsc{CIGALE} \citep{Boquien2019CIGALE}, \textsc{Prospector} \citep{Johnson2021Prospector}, and \textsc{PROSPECT} \citep{Robotham2020PROSPECT}.
Like \textsc{MUFFIT} \citep{DiazGarcia2015MUFFIT} and \textsc{BaySeAGal} \citep{GonzalezDelgado2021}, 
\textsc{AlStar} was developed in the context of J-PLUS and J-PAS and already tested and taylored to handle data from these surveys. A key advantage of \textsc{AlStar} is its ability to simultaneously derive stellar population properties and emission line characteristics.

A detailed description of the code is given in TB23, including applications to integrated and spatially resolved S-PLUS data that are basically identical to those we analyze here. This section briefly reviews the code, its ingredients, and updates. Examples of fits to different regions of M101 are also presented.

\subsection{\alstar}
\label{sec:AlStar}

\alstar\ performs a non-parametric decomposition of a photo-spectrum in terms of a spectral base containing both stellar population and EL elements. The stellar base used in this work comprises composite stellar populations corresponding to periods of constant star formation rate, with 16 $\sim$ logarithmically spaced age intervals spanning from 0 to 14 Gyr, and 3 different metallicities (0.2, 0.5, and $1 Z_\odot$). Note that the adopted metallicity range does not reach the over solar values allowed for in TB23. Our choice of a more restricted $Z$-range is justified by the fact that previous spectroscopic estimates of the stellar metallicity across the body M101 point to values $\lesssim Z_\odot$ \citep{Lin_2013, Simanton-Coogan2017}. Naturally, imposing this external constraint helps reducing the inherent degeneracies of spectral synthesis of stellar populations.

Spectra of these 48 components are built from an updated version of the  \cite{Bruzual_charlot2003} models (see \citealt{2023Martinez-Paredes}). The \cite{Chabrier_2003} initial mass function (IMF) is adopted. This stellar base is expanded to include 94 EL base elements of unitary \Ha\ flux and relative line intensities mimicking those observed in real galaxies, including \hii\ regions, diffuse ionized gas (DIG), and active galactic nuclei.

The empirical EL base of TB23 was updated to include the corresponding nebular continuum emission, which can be significant in spaxels dominated by \hii\ regions (see, e.g., \citealt{2006Corbin, Miranda2025}). The PyNeb tool \citep{2015Luridiana} was used for this purpose. The (relatively minor) dependence of the nebular continuum on the electron temperature is accounted for in an approximate way by following the relation between $T_e$ and gas metallicity in the models by \cite{2017Byler_Neb_cont}, and using the O3N2 $\equiv \log (\oiii/\Hb) / (\nii\Ha)$ calibration of \cite{2013Marino} to estimate the gas metallicity.

The \cite{2000Calzetti} law is used to model dust attenuation. Two dust components are allowed for, one applied to all base components (parametrized by the V-band optical depth $\tau^{\rm ISM}$), and an extra component applied only to populations younger than 10 Myr to represent the dust in the birth clouds ($\tau^{\rm BC})$. The ratio $\tau^{\rm BC} / \tau^{\rm ISM}$ is kept fixed at 1.27, as found in the original \cite{1994calzetti} paper on differential extinction (see also \citealt{Charlot_2000}). 
ELs associated to components compatible with star-formation are attenuated by $\tau^{\rm BC} + \tau^{\rm ISM}$ (as the $\le 10$ Myr components), whereas other components of the EL base are attenuated by just $\tau^{\rm ISM}$.

This general scheme is complemented with an empirical prior which (1) uses a first \alstar\ fit to estimate the combined equivalent width of the \nii$\lambda\lambda$6548,6584  and \Ha\ lines ($W_{\HaNii}$), (2) reduces the EL base to elements where such values of $W_{\HaNii}$ are found in real galaxies, and (3) re-run \alstar. The major effect of this prior is to break the degeneracy between \nii\ and \Ha\ fluxes, which are covered by the same \textit{J0660} narrow band.

The fits are repeated 100 times with fluxes perturbed with gaussian noise of amplitude $\epsilon_\lambda$. Averages over these Monte Carlo (MC) runs are used to define our estimates of physical properties like masses, mean ages, EL fluxes, and others, as well as their uncertainties.

\subsection{Example fits to M101}

\begin{figure}
    \centering
    \includegraphics[width=0.5\textwidth]{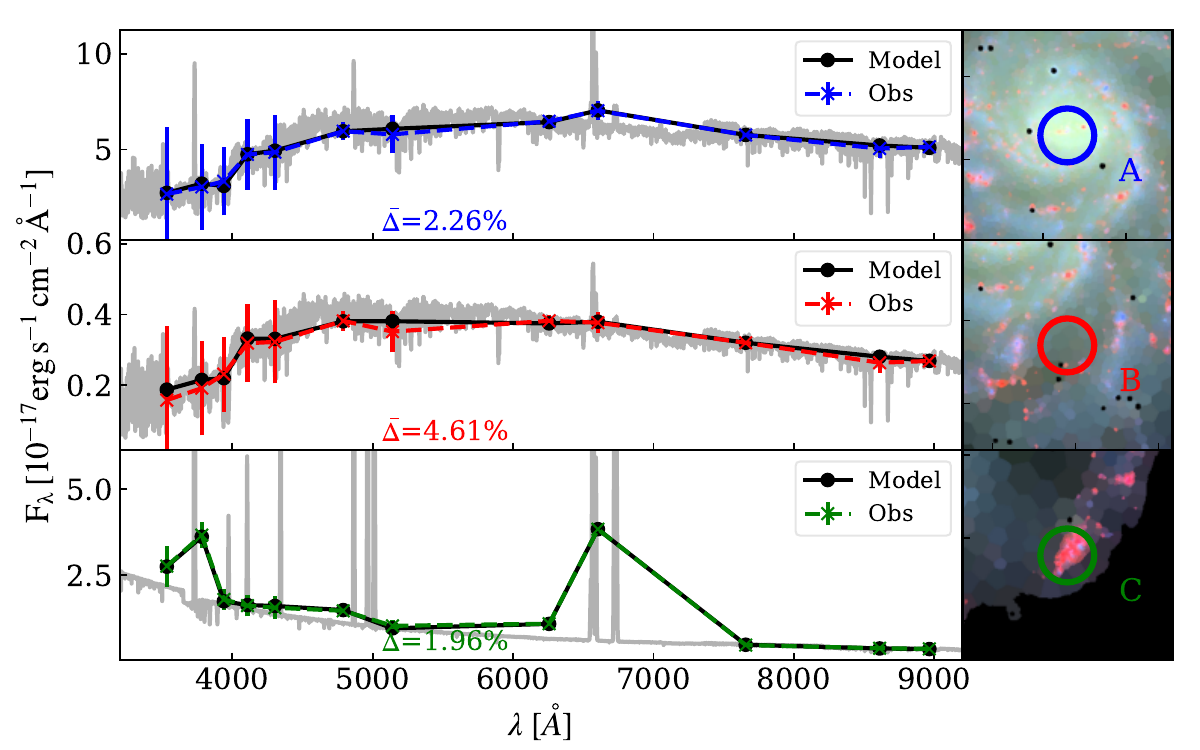}
    \includegraphics[width=0.494\textwidth]{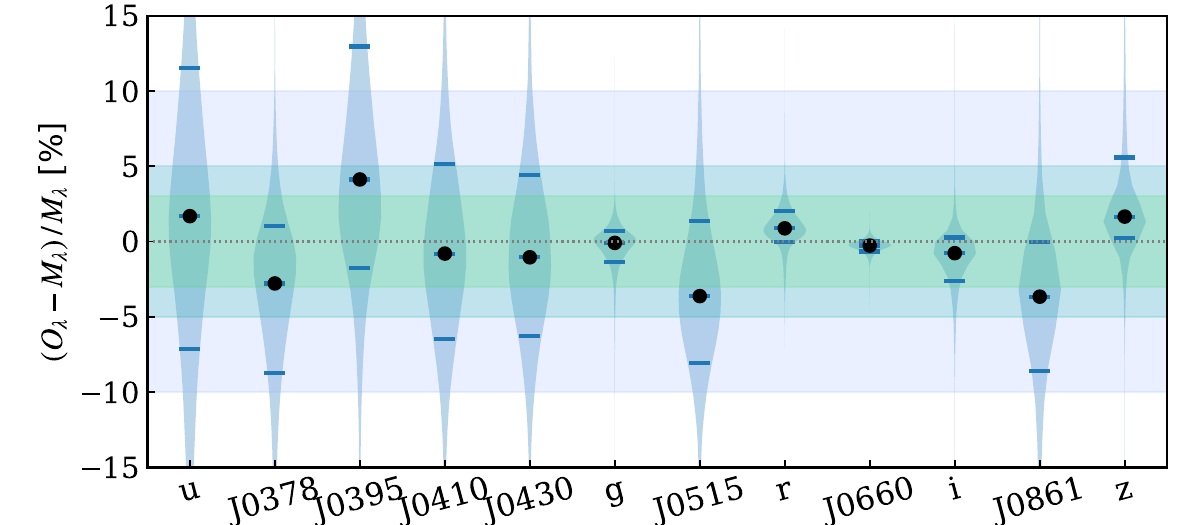}
    \caption{{\em Top:} 
    Example \alstar\ fits for individual spaxels in 3 different regions of M101. 
    Colored lines with error bars show the data ($O_\lambda$), while black lines show the model photometric fluxes ($M_\lambda$), and the gray lines show the corresponding high-resolution model spectrum.
    Images on the right show $278^{\prime\prime} \times 278^{\prime\prime}$ composites built with the \textit{J0660, r}, and \textit{g} fluxes in the R, G, and B channels, respectively.
    {\em Bottom:} Distributions of the $(O_\lambda - M_\lambda) / M_\lambda$ relative residuals of the fits for the 12 J-PLUS bands for 11705 zones in M101. Median residuals are marked by solid black circles, while the horizontal bars mark the 16 and 84 percentiles.
    }
    \label{fig:fit_M101example}
\end{figure}

Fig.\ \ref{fig:fit_M101example} illustrates the quality of our fits. The top panels show three examples of ``J-spectra'', corresponding to the nucleus (A), an interarm zone (B), and an \hii\ region (C), depicted in the stamps on the right. The observed ($O_\lambda$) and model ($M_\lambda$) photo-spectra are shown in the left panels, where the solid black line shows the model photometry while the dashed line and error bars show the data. In all cases, the fluxes shown correspond to those inside a single $1.1^{\prime\prime} \times 1.1^{\prime\prime}$ (re-binned) spaxel, though region B falls inside a Voronoi zone comprising 333 pixels.

As expected from our previous work with S-PLUS (TB23), the \alstar\ fits are very good in all these three examples, which span both different spectral shapes and over an order of magnitude in surface brightness. The mean relative absolute deviation $\overline{\Delta} \equiv \langle |O_\lambda - M_\lambda| / M_\lambda \rangle$ between data and model fluxes is just 2.26, 4.61, and 1.96\% for regions A, B, and C respectively. For the M101 datacube as a whole the median $\overline{\Delta}$ is 3.6\%, and only 323 of the 11706 (2.7$\%$) fits have $\overline{\Delta}> 10\%$. 
These results are also typical of the quality of the fits for NGC 5474, NGC 5585, and NGC 5204, whose median $\overline{\Delta}$ ranges from 2.3 to 3.2\%. For NGC 5477 and UGC8837, the fainter galaxies in the M101 group, the fits are somewhat worse, with median $\overline{\Delta}$ values of 6.0 and 4.0\%, respectively. Fig.\ \ref{fig:fits_adev_extrasM101group} shows example fits for these other galaxies, as well as $\overline{\Delta}$ maps for the whole sample.

The violin plots in the bottom panel of Fig.\ \ref{fig:fit_M101example} show the distributions of the relative residuals $(O_\lambda - M_\lambda) / M_\lambda$ for the 12 bands. As expected, residuals are statistically larger towards the bluer (noisier) bands.

The \alstar\ fits are thus generally very good. From TB23 we further know that the stellar population and EL properties derived from them are reliable, in the sense that they agree with those derived from much more detailed (``\AA\ by \AA'') full spectral fits. 

\section{Results}
\label{sec:Results_M101group}

\begin{figure*}
    \centering
    \includegraphics[width=0.96\linewidth]{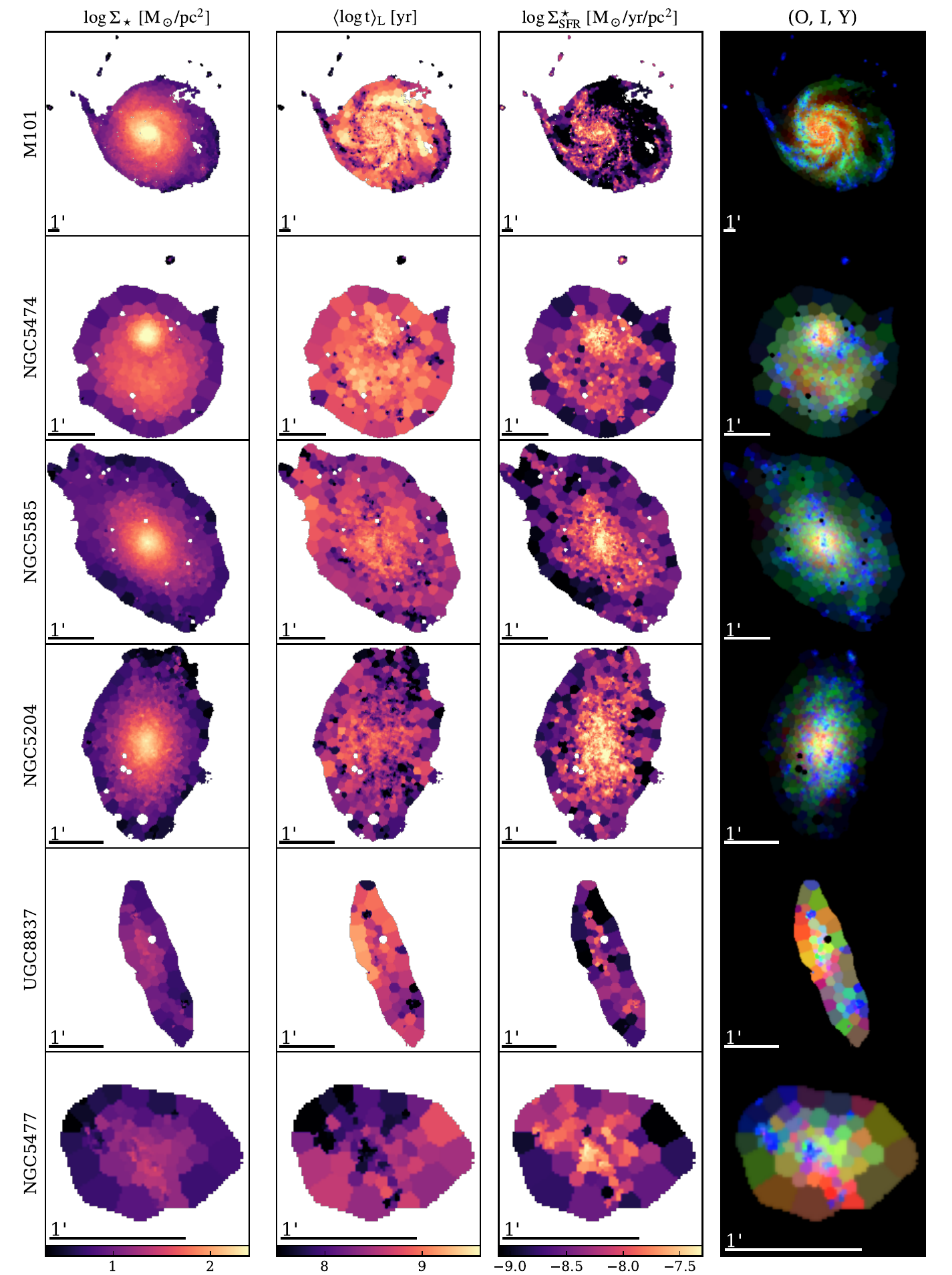}
    \caption{Maps of stellar population properties for galaxies in the M101 group. From left to right: Surface density, mean age, SFR surface density, and an RGB with the fluxes (at 5635 \AA) of Old, Intermediate age, and Young populations. See text for details.
    }
    \label{fig:StellarPropMaps_M101Group}
\end{figure*}

Let us now explore the properties derived from our analysis of J-PLUS data for galaxies in the M101 group. We start with maps of stellar populations (Sec. \ref{sec:StellarPropMaps_M101Group}) and 
EL properties (Sec. \ref{sec:ELPropMaps_M101Group}). Relations between these properties within each galaxy are also investigated (Sec. \ref{sec:age-Z-mass-relations_M101group}).

\subsection{Stellar population maps}
\label{sec:StellarPropMaps_M101Group}

Fig.\ \ref{fig:StellarPropMaps_M101Group} presents maps of stellar population properties obtained with our method for each of the galaxies in the M101 group. The maps are sorted by stellar mass, with larger $M_\star$ at the top.

The first column panels show the stellar mass surface density, $\Sigma_\star$. In M101 (top row)
$\Sigma_\star$ decreases from $\sim 3000 M_\odot\,{\rm pc}^{-2}$ in the nucleus to 
$< 10 M_\odot\,{\rm pc}^{-2}$ in the outer parts. As usual in spirals, the high contrast between arms and interarm regions seen in optical images is not as strong when seen in mass density.
The peak $\Sigma_\star$ values reached by the other galaxies decrease from $\sim 460 M_\odot\,{\rm pc}^{-2}$ for NGC 5474 to $\sim 40 M_\odot\,{\rm pc}^{-2}$ for NGC 5477, following a sequence in $M_\star$.

The second column in Fig.\ \ref{fig:StellarPropMaps_M101Group} shows the luminosity weighted mean log stellar age, defined as $\langle \log t \rangle_L \equiv \sum_j x_j \log t_j$,  where $x_j$ is the fraction of the flux at 5635 \AA\ (our reference normalization wavelength) associated to population $j$ in the base. 
In M101 there is a gradual decrease of $\langle \log t \rangle_L$ towards the outer regions, as well as many ``islands'' of very low age along the arms, corresponding to the many star-forming regions in this galaxy (also detected in our EL maps, as discussed below).
Though a direct quantitative comparison is not warranted because of methodological differences, our negative age gradient is consistent with the trend reported by \cite{2018Hu} on the basis of 
full spectral fitting (their figure 6), as well as with the map produced by \cite{Lin_2013}, whose use broad band photometry from the UV to the mid-IR and a very different modeling strategy based on parametric star formation histories (their figure 11). Both works further identify a young central component, in line with our Fig.~\ref{fig:StellarPropMaps_M101Group}.

This same global description applies to NGC 5574, NGC 5585, and NGC 5204, but this pattern is no longer recognizable in UGC 8837 and NGC 5477, the two less massive galaxies in the group. At the same time, low $\langle \log t \rangle_L$ values become progressively more prevalent as one moves down the $M_\star$ scale.
These results are in line with those obtained from IFS surveys, like the CALIFA-based study by \cite{GonzalezDelgado2014_CALIFA}, who found flatter age gradients for lower mass galaxies, or the MaNGA-based one by \cite{Sanchez_2020_review}, who confirm it for a larger sample.

A detailed study of SFRs will be explored in an upcoming paper, but for the sake of completeness, the third column of Fig.\ \ref{fig:StellarPropMaps_M101Group} shows maps of the SFR surface density derived from the \alstar\ fits, $\Sigma^\star_{\rm SFR}$. These are computed by adding the mass in stars formed in the last $t_{\rm SF}$ years and dividing it by $t_{\rm SF}$ (as in \citealt{2007Asari}). We adopt $t_{\rm SF} = 100$ Myr as a compromise between the desire to map the recent SFR and the need to group base-ages to produce a more robust estimator (see, e.g., the experiments in \citealt{2004CidFernandes_Gu}). The maps combine the patterns of the first and second panels,  
showing higher $\Sigma^\star_{\rm SFR}$ in regions of higher $\Sigma_\star$, as expected from the star-forming main sequence (MS, e.g. \citealt{2020Enia}), but also in regions of low $\langle \log t \rangle_L$.
We also note the strong asymmetry in the $\Sigma^\star_{\rm SFR}$ map of M101.

Finally, the rightmost images in Fig.\ \ref{fig:StellarPropMaps_M101Group} 
combine the fluxes at 5635 \AA\ of Young ($t \le 10$ Myr), Intermediate age (10 Myr $< t < 1$ Gyr), and Old ($t \ge 1$ Gyr) populations into an RGB composite. The overall inside-out R $\rightarrow$ G $\rightarrow$ B run of colors in the M101 to NGC 5204 panels reflects the negative mean age gradient previously seen in the $\langle \log t \rangle_L$ maps, while for UGC 8837 and NGC 5477 no organized pattern is discernible. In M101, where the spiral arms are well defined, old populations are also ubiquitous in the inter-arm regions, while the arms themselves are dominated by intermediate age and young populations.

Maps of the dust optical depth, along with dereddened color composites (similar to those presented in TB23 for NGC1365), are presented in the appendix (Fig.\ \ref{fig:dust_all_gals_M101group}). Unlike in the M51 group (Thainá-Batista et.\ al., in prep.), dust attenuation is generally small throughout all galaxies in the M101 group (typically around 0.2 mag in terms of $A_V$, the most notable exceptions occurring in star-forming regions, which are dustier. $A_V \sim 0.2$ mag is also the typical value obtained by \citealt{2015GonzalezDelgadoR} for Sd galaxies, which can be seen as a typical morphological type for galaxies in the M101 group (Tab. \ref{tab:sample}).

The uncertainties in properties discussed above are estimated from the dispersion among the MC realizations performed by \alstar.  Using the robust $\sigma_{\rm NMAD}$\footnote{$\sigma_{\rm NMAD}(x) = 1.4826 \times {\rm median}( | x - {\rm median}( x ) | )$.} statistic to quantify these noise-related uncertainties, we find median uncertainties of 0.22--0.25 dex in $\log \Sigma_\star$ over the six M101 group galaxies, rising to $\sim 0.4$ dex in regions of strong star formation. Mean light‐weighted ages have median $\sigma_{\rm NMAD}(\langle\log t\rangle_L)$ values of $0.3$--0.5 dex, with larger uncertainty for ages near $10^8$ yr and in the fainter outskirts. Stellar metallicities exhibit smaller formal errors ($\sim0.2$ dex), largely reflecting the restricted metallicity grid used, while SFR surface densities have $\sigma_{\rm NMAD}(\log \Sigma_{\rm SFR}) \sim 0.2$ dex when considering only zones where young populations contribute at least 30\% of the 5635 \AA\ flux. Sec.\ \ref{sec:Uncertainties} in the appendix discusses these and other uncertainties in more detail.

\subsection{Emission line maps}
\label{sec:ELPropMaps_M101Group}

\begin{figure*}
    \centering
    \includegraphics[width=0.93\linewidth]{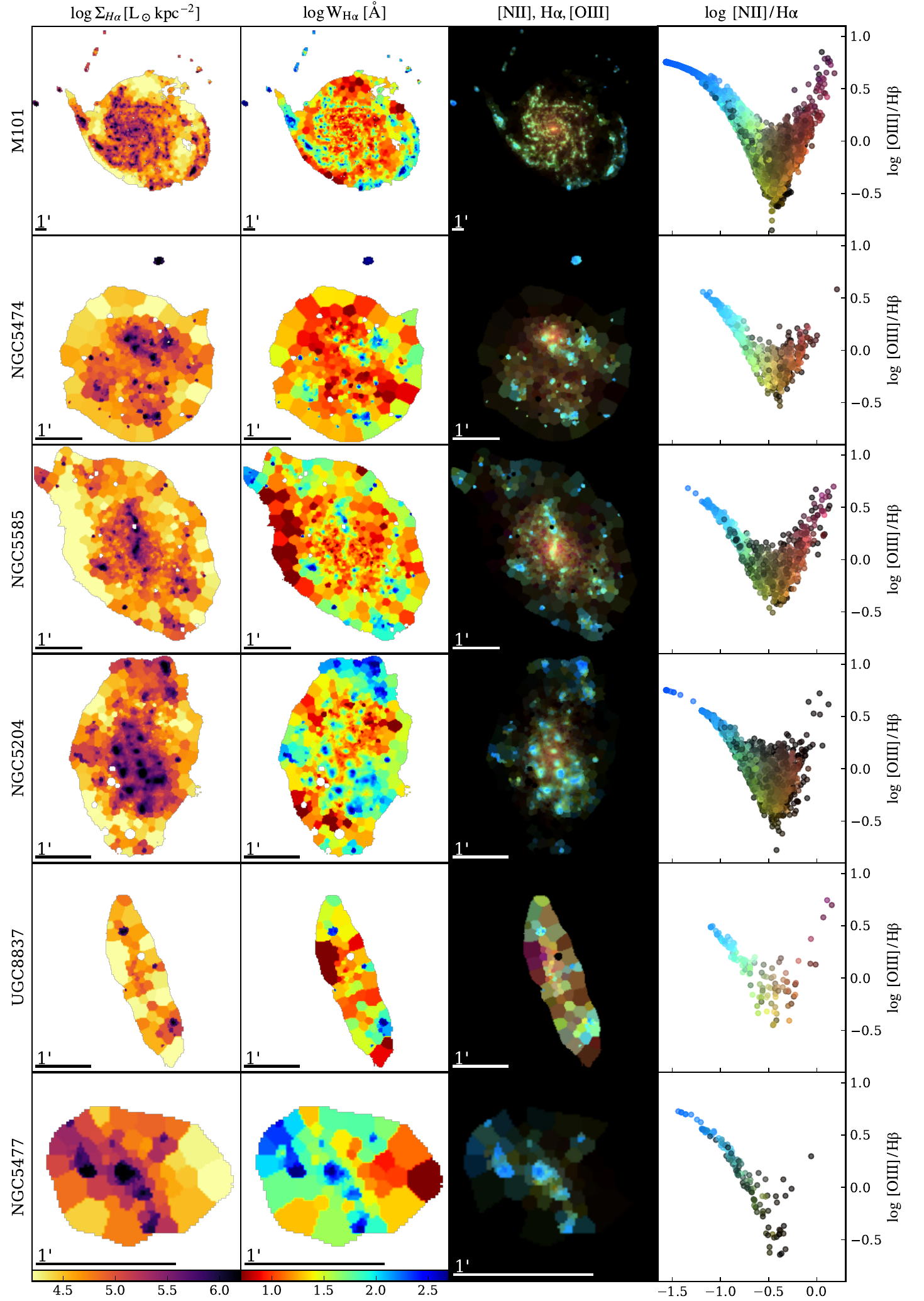}
    \caption{Maps of emission line properties. From left to right: \Ha\ surface brightness,
    \Ha\ equivalent width,  RGB with the (\nii, \Ha, \oiii) fluxes, BPT diagram of the spaxels of each galaxy, with points colored as in the RGB.}
    \label{fig:ELPropMaps_M101Group}
\end{figure*}

Let us now examine the EL properties derived from our fits. This is done in Fig.\ \ref{fig:ELPropMaps_M101Group}, whose first column shows the \Ha\ surface brightness ($\Sigma_{\Ha}$) maps for galaxies in the M101 group, ordered as in Fig.\ \ref{fig:StellarPropMaps_M101Group}. Star-forming regions stand out clearly in all cases. Diffuse emission is also detected, particularly in the central regions of the top four galaxies and in between the arms of M101. Though unsurprising in physical terms, it is interesting to note the strong correspondence between these \Ha\ maps and the young stellar populations mapped in Fig.\ \ref{fig:StellarPropMaps_M101Group}, which shows that our methodology is able to identify young populations by their contributions to both ELs and stellar continuum.

The second column in Fig.\ \ref{fig:ELPropMaps_M101Group} shows maps of the \Ha\ equivalent width ($W_{\Ha}$). 
The maps span values from as low as 1 \AA\ to over 1000 \AA.
Again, one sees an excellent correspondence between regions of large $W_{\Ha}$ and those of small $\langle \log t \rangle_L$ in Fig.\ \ref{fig:StellarPropMaps_M101Group}.
Regions painted in red-orange colors, where $W_{\Ha}$ values are relatively small, are visibly more diffuse than those in blue ($W_{\Ha} > 100$ \AA), which appear knotty. This diffuse emission matches the characteristics of the mixed-DIG component as defined by \cite{2018_Lacerda}, with \WHa\ in the $\sim 3$--14 \AA\ range.
Given the widespread star-formation in these galaxies, it is likely that this diffuse emission is mainly powered by leakage of ionizing photons from \hii\ regions \citep{2024Watkins}.

Regions B and C of Fig.\ \ref{fig:fit_M101example} are examples of regions whose ELs are dominated by DIG and \hii\ regions, respectively. Their $W_{\Ha}$ values are 9.4 and 700 \AA, respectively, while 
$\nii/\Ha = 0.5$ and 0.10. The large $W_{\Ha}$ in region C is in fact evident from its J-spectrum in Fig.\ \ref{fig:fit_M101example}, which also shows a strong \oii\ ($W_{\oii} = 231$ \AA\ according to our fits). The nucleus (region A), has intermediate values, with $W_{\Ha} = 15.9$ \AA\ and $\nii/\Ha = 0.37$.

The third column of Fig.\ \ref{fig:ELPropMaps_M101Group} combines the \nii, \Ha, and \oiii\ maps onto an RGB composite which synthesizes our main results regarding ELs and their spatial variations. 
The fourth column shows the corresponding BPT diagrams (\citealt{1981BaldwinPhiTerl}), $\log \oiii/\Hb$ vs.\ $\log \nii/\Ha$, where points are colored as they appear in the RGB panel. The rather schematic look of our BPT diagrams stems from our very methodology to estimate ELs out of J-PLUS photometry, which involves a discrete EL base built on the basis of the BPT diagram itself (see TB23).

The redder hue in the central regions of the top four galaxies, as well as in inter-arm regions of M101, trace regions of elevated $\nii/\Ha$, characteristic of DIG-like emission.
Star-forming regions gradually change their colors in this image from green in the inner disk to blue in the outer parts due to a systematic increase of the ratio between \oiii\ (plotted in the B channel) to \Ha\ (G). 
In the case of M101 we associate this behavior with its well-documented negative nebular metallicity gradient (e.g., \citealt{2007Bresolin}, \citealt{2016Croxall}). This is confirmed in its BPT diagram, where spaxels with bluer points of the outer \hii\ regions in the RGB composite populate locii in the upper-left, low O/H region of the BPT. The diagram also shows how reddish regions in the (\nii, \Ha, \oiii) composite map onto the right wing of the BPT, as one would expect for DIG-like ELs. Note also how dark (low flux, mostly inter-arm) regions in the EL RGB map onto the right wing of the BPT.
The same trend seems also present in the (\nii, \Ha, \oiii) images of NGC 5474, NGC 5585, and NGC 5204, although not as marked as in M101.

Both our $W_{\Ha}$ map, BPT diagram, and radial trends in line ratios for M101 match well the results of \cite{2018Hu}. We note, however, that this conclusion is based on a coarse visual analysis, as their ``maps'' are not continuous, but focused on individual \hii\ regions, which also explains why they hardly sample the right wing, DIG-dominated regions of the BPT in M101.

Typical (MC-based) uncertainties in $\Sigma_{\Ha}$ range from $0.1$ to $0.3$ dex across our galaxies, with the larger relative uncertainties occurring in regions of weak line emission. Restricting the analysis to regions where $W_{\Ha} > 30 \AA$ we obtain $\sigma_{\mathrm{NMAD}}(\log \Sigma_{\Ha})$ ranging from 0.04 to 0.13 dex.
For $W_{\Ha}$ we obtain $\sigma_{\mathrm{NMAD}}(\log W_{\Ha})\sim0.1$ dex for most galaxies (up to 0.3 dex in the noisiest cases).
Ratios such as $\nii/\Ha$ show median uncertainties near 0.2 dex, For a lengthier discussion of uncertainties in EL properties, see Sec.\ \ref{sec:Uncertainties} in the appendix.

\subsection{Age-metallicity-mass relations}
\label{sec:age-Z-mass-relations_M101group}

\begin{figure*}
    \centering
    \includegraphics[width=\linewidth]{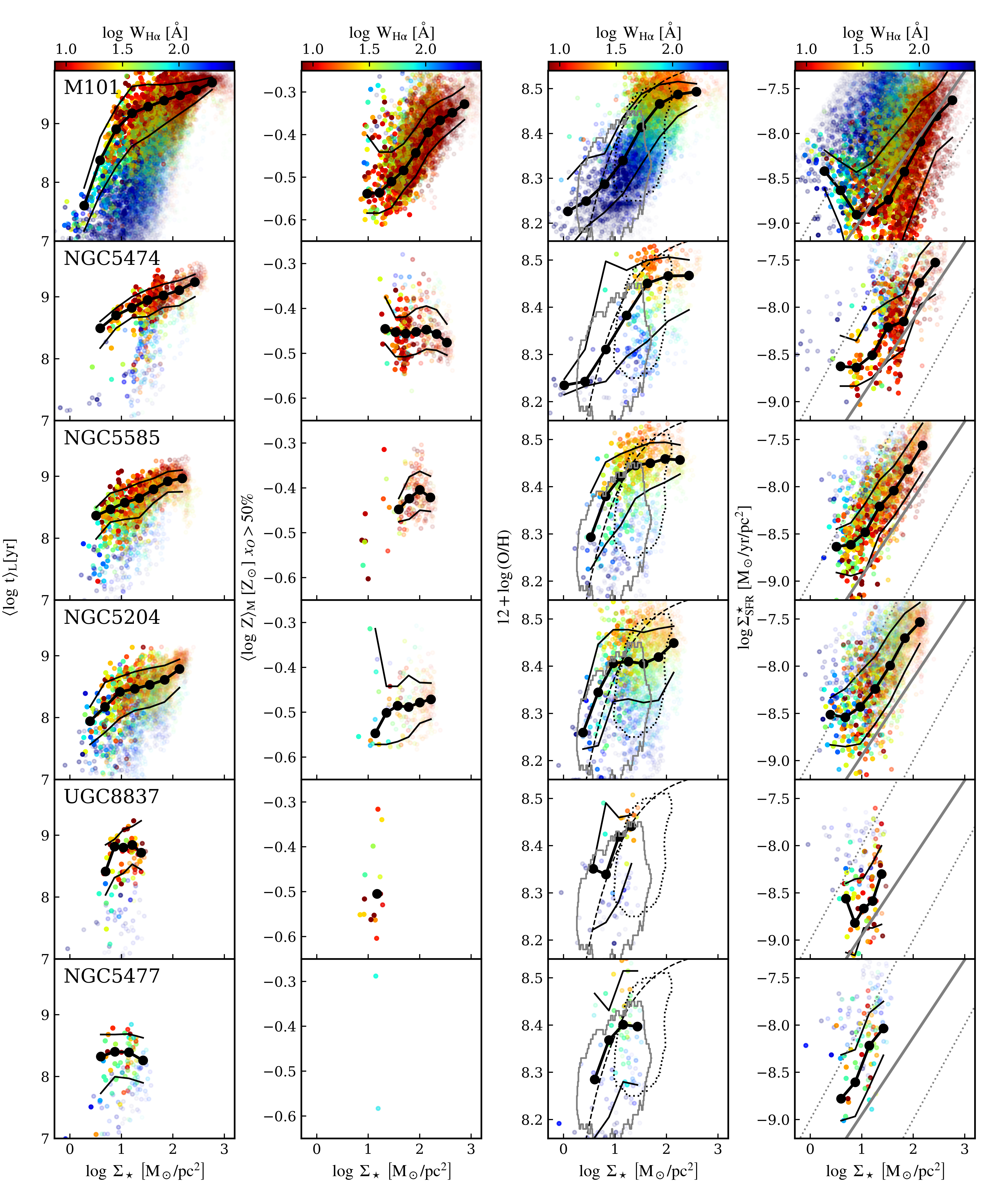}
    \caption{Scaling relations for galaxies in the M101 group. Individual zones are plotted as circles colored by $\log W_{\Ha}$. The thick black lines with circle marks the median curves for bins in $\log \Sigma_\star$, while thin lines show the corresponding 16 and 84 percentile curves.
    In the third column, the black dashed line is the median curve from \cite{2016Barrera-Ballesteros} for 653 MaNGA galaxies, the gray line shows the 80\% contour of  \cite{2016Barrera-Ballesteros} for MaNGA galaxies with $\log M_\star/M_\odot < 9.2 $, while the dotted contour indicates the typical locii of Sd galaxies (from \citealt{Sanchez_2020_review}). Dashed lines in the fourth column panels mark lines of specific SFR $=  0.01$ and 1 Gyr$^{-1}$, while the solid gray line is the relation obtained by \cite{2020Enia} --- see text for details.
    }
    \label{fig:age-Z-Mass_M101group}
\end{figure*}

Let us now examine our results in terms of the relations among the derived properties. Some of the most studied relations in galaxies are the mass-age, mass-metallicity, and mass-SFR relations. These relations were mapped both for galaxies as a whole (e.g., \citealt{Gallazzi2005, Renzini2015}), and with spatially resolved observations (e.g., \citealt{2016GonzalezDelgadoR, Sanchez_2020_review}).

Fig.\ \ref{fig:age-Z-Mass_M101group} plots the mean stellar age and metallicity as well as the nebular metallicity and SFR surface density against the local stellar mass surface density.
As in previous figures, galaxies are sorted by mass. Points are colored by $\log W_{\Ha}$, which facilitates the identification of spaxels associated to star-forming regions. 
Bins along the $\Sigma_\star$-axis were defined to draw median curves (thick line with circles) as well as the 16 and 84 percentiles (gray lines).
Given the systematic decrease of $\Sigma_\star$ with distance to the center, these plots can be seen as equivalent to radial profiles, with the advantage that one circumvents the need to define a meaningful radial coordinate in these distorted systems.

\subsubsection{The mass-age relation}

The tendency for more massive galaxies to have older stellar populations (a.k.a.\ ``downsizing''; \citealt{PerezGonzalez2008}), has a local counterpart within galaxies \citep{GonzalezDelgado2014MZRletter}. The left panels in Fig.\ \ref{fig:age-Z-Mass_M101group} show how our $\langle \log t \rangle_L$ age indicator varies as a function of $\Sigma_\star$ for galaxies in the M101 group.

Except for the dwarfs, all other galaxies in the group show a clear mass-age relation, with relatively tight ranges in $\langle \log t \rangle_L$ for fixed $\Sigma_\star$. The growth of age with mass depicted in these plots reflects the overall negative age gradients seen in Fig.\ \ref{fig:StellarPropMaps_M101Group} and well documented in the literature, particularly that based on IFS work (\citealt{Sanchez_2020_review} and references therein). The two galaxies that deviate from this pattern are UGC 8837 and NGC 5477, the smallest and least massive galaxies of the group, which exhibit a flat mass-age relation, in line with the IFS-based results of \cite{2015GonzalezDelgadoR}, which point to flatter age profiles at lower galactic masses.

Notably, regions deviating from the median trend do so mostly towards younger ages and have larger $W_{\Ha}$ (over 50 \AA), as indicated by their green-blue symbols. Despite the abundance and exuberance of these star-forming regions, they occupy but a fraction of the area of the galaxies, and thus have relatively little influence over the median relation. We note in passing that the systematic gradient in symbol colors along the vertical direction in these panels reinforces the connection between  $\langle \log t \rangle_L$ and $W_{\Ha}$ noted while comparing their maps in Figs.\ \ref{fig:StellarPropMaps_M101Group} and \ref{fig:ELPropMaps_M101Group}.

\subsubsection{The stellar mass-metallicty relation}

To investigate the stellar mass-metallicity relation (MZR) we use the mass weighted mean (log) stellar metallicity, $\langle \log Z \rangle_M \equiv \sum_j \mu_j \log Z_j$, where $\mu_j$ denotes the mass fraction associated to component $j$.
Since most of the mass in a stellar population lies in low mass, long lived stars, this index naturally gives a larger weight to old populations. This is in fact desirable, given the notorious difficulty in estimating stellar metallicities in young populations (e.g., \citealt{Cid2005}). 
Still, given the profusion of star-forming regions in our galaxies, 
many spaxels have their light overwhelmed by young stars to the point of affecting the derived $\langle \log Z \rangle_M$ values. 
To mitigate this effect, we focus on regions where populations older than 1 Gyr account for more than $x_O = 50\%$ of the flux at 5635 \AA.

The second column of Fig.\ \ref{fig:age-Z-Mass_M101group} shows the resulting stellar MZRs. Again, points are colored by $W_{\Ha}$, and median, 16, and 84 percentile relations are indicated. Because of its much larger statistics, M101 exhibits a better defined stellar MZR, with a clear increase of $\langle \log Z \rangle_M$ for increasing $\Sigma_\star$.
Ultimately, this reflects a negative stellar metallicity gradient, with inner (denser) regions being more metal rich ($\sim 1/2$ solar), than the outer (less dense) ones ($\sim 1/4$ solar at the median point). The coloring scheme shows that the scatter in the MZR is related to $W_{H\alpha}$ and hence (indirectly) to $\langle \log t \rangle_L$, in the sense that, for a fixed $\Sigma_\star$, younger zones have larger metallicities. This is consistent with chemical evolution, but one should recall the caveats about estimates of the stellar metallicity in the presence of young populations.

Unlike M101, NGC 5474 exhibits a flat MZR. The values of $\langle \log Z \rangle_M$ are generally smaller than those in M101, oscillating around 0.3 solar. 
NGC 5585 and NGC 5204 exhibit, at best,  weak MZRs.
Nothing can be said with our data about the stellar MZR in the two smallest galaxies in the sample, as they have too few regions satisfying our $x_O > 50\%$ criterion to obtain a statistically meaningful estimate of $\langle \log Z \rangle_M$.

\subsubsection{The nebular mass-metallicity relation}

The nebular version of the local MZR is shown in the third column of Fig.\ \ref{fig:age-Z-Mass_M101group}. We estimate $12 + \log {\rm O/H}$ by means of the O3N2 index,
following the  calibration by \cite{2013Marino}. Since this strong-line method is only applicable to star-forming regions, we exclude all zones whose ELs place them above the \cite{2003KalffmannSDSS} line in the BPT diagram. We further exclude zones where $W_{\Ha} < 14$ \AA\ to mitigate effects of the mixed-DIG component discussed by \cite{2018_Lacerda}, although this extra constraint has negligible effects on the resulting MZRs. Finally, zones larger than 50 pixels are disconsidered in order to avoid their large weight on the overall statistics as well as to minimize the mixing of very different spatial scales.

The correlation between $\Sigma_\star$ and O/H is evident in M101, as reinforced by the thick black solid lines in Fig.\ \ref{fig:age-Z-Mass_M101group}, which trace the median relation. 
This nebular MZR is in line with the well documented O/H gradient in M101 (e.g., \citealt{2003Kennicutt, 2020Berg}), as well as with MZRs derived from IFS studies of other galaxies. The points overlap well with the dotted figure, which outlines the 50\% contour for Sd galaxies from the \cite{Sanchez_2020_review} review (shifted by $-0.26$ dex to correct for differences in the IMF).
The median curve is also close to the MZR obtained by \cite{2016Barrera-Ballesteros} in their analysis of 653 MaNGA galaxies, marked with a dashed line (also adjusted for IMF), a difference which we attribute to the contribution of more massive, metal rich sys
in their sample, particularly at large $\Sigma_\star$.

Nebular MZRs are also obtained for NGC 5474, NGC 5585, and NGC 5204, albeit not as well defined as in M101. 
An MZR is only weakly suggested in the cases of  UGC 8837 nor in  NGC 5477, in line with IFS-based studies, which generally find weak or no systematic nebular metallicity gradients in low mass galaxies (e.g., \citealt{2017Belfiore}).
Our results for these five less massive galaxies in the M101 group overlap well with the region occupied by spaxels in MaNGA galaxies with $\log M_\star/M_\odot < 9.2$ from \cite{2016Barrera-Ballesteros}, drawn as a solid gray line (also adjusted for IMF). These contours also overlap with the outer regions of M101.

Evidently, these results are subject to the usual caveats and limitations of strong-line methods, further compounded by the uncertainties inherent in our original approach for estimating EL fluxes from purely photometric data.

\subsubsection{The spatially resolved star-forming main sequence}

The rightmost panels of Fig.\ \ref{fig:age-Z-Mass_M101group} show the relation between our estimated SFR surface density derived from the \alstar\ fits, $\Sigma^\star_{\rm SFR}$, with $\Sigma_\star$. Besides the median and percentile curves, these panels show lines of specific SFR of 0.01 and 1 Gyr$^{-1}$, as well as the relation obtained by \cite{2020Enia} on the basis of multi-band photometry from the UV to the far-IR of eight $M_\star = 1$--$5 \times 10^{10} M_\odot$ grand design spirals (thick gray line).

The effects of spatial sampling (which varies from 39 pc spaxels to 2 kpc Voronoi zones in our galaxies) on the scatter in this relation are explored in detail in a separate study, but we anticipate that they are large. The median curves, however, should be immune to this issue, so let us focus on them.

All galaxies exhibit  $\Sigma^\star_{\rm SFR}$-$\Sigma_\star$ relations either $\sim$ parallel to the \cite{2020Enia} line, or, in the case of M101, very close to it, at least for $\Sigma_\star \gtrsim 10 M_\odot/{\rm pc}^2$. In all cases, regions of large $W_{\Ha}$ lie systematically above the median relations, consistent with the fact that $W_{\Ha}$ is itself a tracer of specific SFR. The flattening or upturn seen at low $\Sigma_\star$ is in fact not surprising considering that most zones contributing to the median curve in these regions have large $W_{\Ha}$, as denoted by the green-blue colors they are painted with.
This is likely a selection effect, in the sense that a low $\Sigma_\star$ region will only make it into the galaxy mask if it is bright enough, a criterion that which naturally favors star-forming regions over more passive ones. 
More interestingly, the right panels in Fig.\ \ref{fig:age-Z-Mass_M101group} further show a systematic upward shift in the median relation as one moves down the panels, i.e., as $M_\star$ decreases, or to later morphological type, as observed in CALIFA  \citep{2016GonzalezDelgadoR, 2017GonzalezDelgadoR}.

We close by noticing that our values for the total stellar mass ($2.2\times 10^{10} M_\odot$) and SFR ($3.2 M_\odot/{\rm yr}$) derived from the spatially resolved data for M101 cube agree within 6\% and 24\% respectively with the estimates by \cite{2020Enia} using UV to far IR data. This reinforces the conclusion of TB23 that, despite the limitations of our data and somewhat unorthodox aspects of our analysis (mainly the EL-base), our methodology yields physical properties in line with those derived from more conventional methods.

\section{Discussion}
\label{sec:Discussion}

\begin{figure*}
    \centering
    \includegraphics[width=\linewidth]{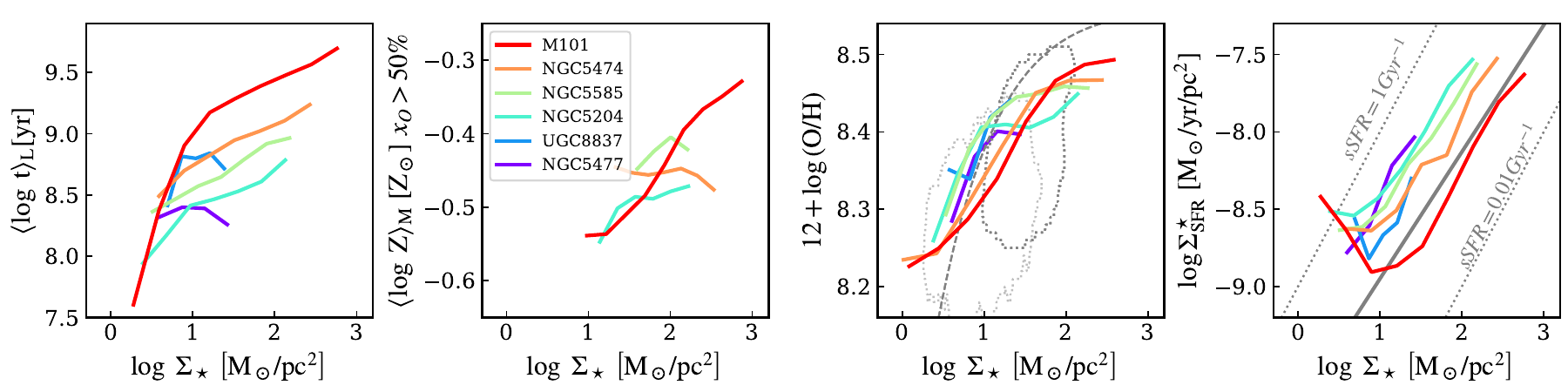}
    \caption{Median scaling relations for all galaxies.
    The dotted lines in the right panel represent lines of specific SFR of 0.01 and 1 Gyr$^{-1}$, while the thick gray line shows the star-forming MS relation obtained by \cite{2020Enia} for nearby spirals.
    }
    \label{fig:MedianRelations4M101group}
\end{figure*}

In this section we compare the scaling relations and other derived properties of the galaxies in the M101 group. We explore similarities and differences between the galaxies and compare our findings with results from the literature.

In order to facilitate the inter-comparison of the scaling relations among the different galaxies in the group, Fig.\ \ref{fig:MedianRelations4M101group} overplots the median relations in Fig.\ \ref{fig:age-Z-Mass_M101group} for all six galaxies.
The local mass-age relation ($\Sigma_\star$-$\langle \log t \rangle_L$) shows that all galaxies $\sim$ converge at \(\Sigma_\star \lesssim 10\ M_\odot/\text{pc}^2\) and diverge at higher densities, with more massive ones generally exhibiting older stellar populations for the same \(\Sigma_\star\) \citep{PerezGonzalez2008, 2015GonzalezDelgadoR}. This trend is consistent with the expected inside-out growth of galaxies, where central regions are more evolved. 
Only UGC 8837 spoils this sequence.

The stellar and nebular MZRs exhibit distinct behaviors. The stellar MZR in the second panel of Fig. \ref{fig:MedianRelations4M101group} shows that our most massive galaxy, M101, reaches the highest $\langle \log Z/Z_\odot \rangle_M$ values, with a clear gradient towards denser regions. In contrast, NGC 5474, NGC 5585 and NGC 5204 exhibit flatter relations, suggesting a less efficient metal enrichment process, possibly due to the shallower gravitational potential wells in these low mass systems. This is in line with results from IFS-based studies \citep{Neumann2021}. Beyond this, the values of $\langle \log Z/Z_\odot \rangle_M$ are around -0.5, which is consistent with Sd galaxies  \citep{2015GonzalezDelgadoR}.

The nebular MZR reveals systematic trends with \(\Sigma_\star\) in all galaxies. In M101, O/H increases $\sim$ steadily with \(\Sigma_\star\). This gradient becomes less pronounced in less massive galaxies like NGC 5474, reflecting lower overall enrichment levels. This is consistent with \cite{2017Belfiore}, which observes flatter radial gradients of O/H as $M_\star$ decreases.

The spatially resolved $\Sigma^\star_{\mathrm{SFR}}$-$\Sigma_\star$ relations for all galaxies align well with the MS of star formation. For M101, the relation matches the \cite{2020Enia} results at $\Sigma_\star \gtrsim 10 M_\odot \mathrm{pc}^2$, while less massive galaxies show systematic $M_\star$-dependent offsets, possibly reflecting differences in star formation efficiency and evolutionary states. The upturn towards large specific SFR at low $\Sigma_\star$ occurs because of the star-forming regions sampled in the outskirts of our galaxies, whose $W_{\Ha}$ values exceed 200 \AA.

Finally, we note that the widespread star-formation across all galaxies in the group indicates that they have not undergone any substantial environmental quenching. 
The efficacy of many environmental processes is known to increase with the mass of the group or cluster (e.g., \citealt{Alonso2012}; \citealt{2019Raj}).
In terms of its total stellar mass, $2.5 \times 10^{10} M_\odot$ according to our analysis, M101 is a relatively low mass group, which may be the reason why its galaxies show no sign of suppressed star-formation.
This is in line with the results of \cite{GonzalezDelgado2022} on the effect of the stellar mass of the group on the excess of quenched galaxies with respect to the field. They find that, for a fixed $M_\star$, the quenched fraction excess in low mass groups (defined as those with $<5 \times 10^{11}$~$M_\odot$ in stars, the regime of the M101 group) is significantly lower than in more massive structures.

Clearly, at this point, these are just plausible speculations, based on results of previous integrated-light work. As we extend the work on the 2D characterization of galaxy properties to other groups and clusters, we will be better equipped to evaluate environmental effects on the properties of galaxies at local scales, as reported by other works (see e.g. \citealt{Coenda2019}, \citealt{Bluck2020b}, \citealt{Epinat2024} ).

\section{Summary}
\label{sec:conclusions}

This paper inaugurates a series of studies of galaxies in nearby groups and clusters based on the 12-bands optical imaging photometry data from the J- and S-PLUS surveys. 
This first article presented our data-handling and photo-spectral synthesis methodologies, as well as results for the M101 group derived from J-PLUS observations.
We have presented (Figs.\ \ref{fig:StellarPropMaps_M101Group}--\ref{fig:age-Z-Mass_M101group}):

   \begin{enumerate}
      \item Maps of the stellar mass surface density ($\Sigma_\star$), mean stellar age ($\langle \log t \rangle_L$), SFR surface density derived from the synthesis ($\Sigma^\star_{\rm SFR}$), maps of the stellar populations divided into young, intermediate, and old age bins.
      
      \item Maps of emission line properties: \Ha\ fluxes, equivalent width, and (\nii, \Ha, \oiii) RGB composites, along with the corresponding BPT diagram.
      
      \item The (spatially resolved) relations between the mean stellar age, stellar and nebular metallicities, and $\Sigma^\star_{\rm SFR}$  with $\Sigma_\star$.   
   \end{enumerate}

We have also compared the local scaling relations for the different galaxies in the group, obtaining (a) $\Sigma_\star$-age relations shifted towards younger ages as $M_\star$ decreases; (b) a well defined stellar MZR in M101, but flatter or undefined relations of the less massive galaxies; (c) a nebular MZR in line with that found in IFS studies of galaxies in the same mass range; (d)  $\Sigma^\star_{\mathrm{SFR}}$-$\Sigma_\star$ MS relations offset towards larger specific SFR levels as $M_\star$ decreases.

We have thus reached our aim of characterizing the spatially resolved stellar population and EL properties of galaxies in the M101 group. Our results underscore the power of J-PLUS to perform spatially resolved analyses of galaxy properties, providing insights comparable to those obtained from traditional IFS surveys.

By itself, however, this study is not enough to assess the kind and intensity of  environmental effects acting upon galaxies in groups. To do so, one must first assemble similar characterization studies for galaxies in different environments, and this study represents a first step in this direction.
The same overall methodology employed here will be applied to other nearby groups and clusters in future communications, with the ultimate goal of building a large and homogeneous set of IFS-like observational data and analysis products on galaxies under different  environments, enabling meaningful comparative studies of systems of different masses, dynamical and evolutionary states.

\begin{acknowledgements}
    This work was supported by CAPES under grant 88881.892595/2023-01 and FAPESC (CP 48/2021). 
    RCF acknowledges support from CNPq (grants 302270/2018-3 and 404238/2021-1).
    JTB, RGD, RGB, JRM, GMS and LADG acknowledge financial support from the Severo Ochoa grant CEX2021-001131-S funded by MCIN/AEI/ 10.13039/501100011033 and to grant PID2022-141755NB-I00.
Based on observations made with the JAST80 telescope and T80Cam camera at the Observatorio Astrofísico de Javalambre (OAJ), in Teruel, owned, managed, and operated by the Centro de Estudios de Física del Cosmos de Aragón. We acknowledge the OAJ Data Processing and Archiving Department (DPAD) for reducing and calibrating the OAJ data used in this work, as well as the distribution of the data products through a dedicated web portal. Funding for OAJ, UPAD, and CEFCA has been provided by the Governments of Spain and Aragón through the Fondo de Inversiones de Teruel and their general budgets; the Aragonese Government through the Research Groups E96, E103, E16\_17R, E16\_20R and E16\_23R; the Spanish Ministry of Science and Innovation (MCIN/AEI/10.13039/501100011033 y FEDER, Una manera de hacer Europa) with grants PID2021-124918NB-C41, PID2021-124918NB-C42, PID2021-124918NA-C43, and PID2021-124918NB-C44; the Spanish Ministry of Science, Innovation and Universities (MCIU/AEI/FEDER, UE) with grant PGC2018-097585-B-C21; the Spanish Ministry of Economy and Competitiveness (MINECO) under AYA2015-66211-C2-1-P, AYA2015-66211-C2-2, AYA2012-30789, and ICTS-2009-14; and European FEDER funding (FCDD10-4E-867, FCDD13-4E-2685). The Brazilian agencies FINEP, FAPESP, and the National Observatory of Brazil have also contributed to J-PLUS. We also thank the comments from J. A. Fernández-Ontiveros, P.T. Rahna, and E. Telles.
\end{acknowledgements}

\bibliographystyle{aa}
\bibliography{aa54704-25}

\begin{appendix}

\section{Complementary material}

\subsection{Pre-processing}
\begin{figure*}
    \centering
    \includegraphics[width=\linewidth]{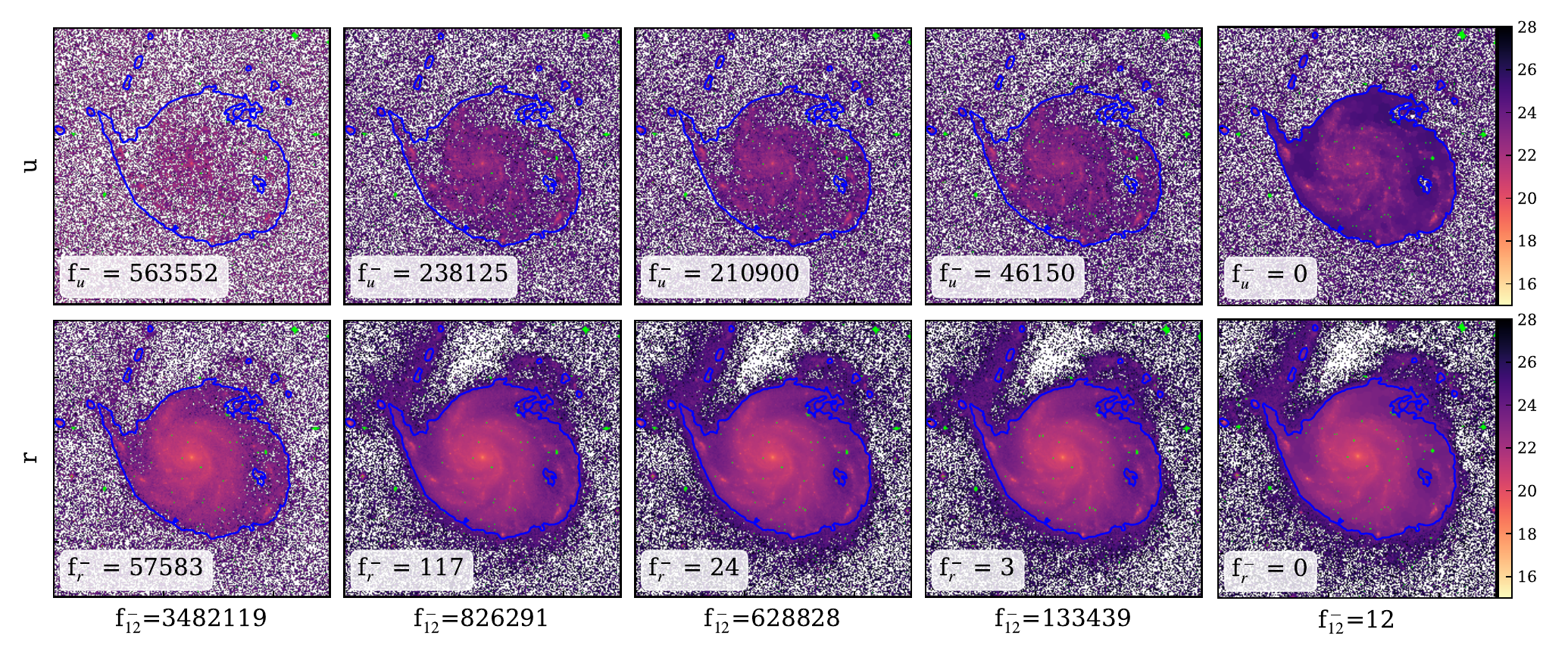}
    \caption{
    Step-by-step of the pre-processing applied to the J-PLUS data cube of M101. The columns show, respectively: the original data, Butterworth filter, PSF homogenization, $2 \times 2$ binning, and Voronoi binning (defined by the \textit{u}-band). The blue line represents the contour of the galaxy mask. The first and second rows are the surface brightness maps (in AB mag/arcsec$^2$) in the \textit{u} and \textit{r}-band respectively. Pixels with $< 0$ flux appear in white. The legends f$_{u}^-$ and f$_{r}^-$ show the number of pixels with negative fluxes inside the galaxy mask in the u and \textit{r} bands, while the f$^-_{12}$ numbers under the plots are the sum of the $f_\lambda < 0$ pixels in all the 12 bands for each of the pre-processing steps. Masked stars are marked in green.}
    \label{fig:pre-processing_steps_M101}
\end{figure*}

\begin{figure*}
    \centering
    \includegraphics[width=\textwidth]{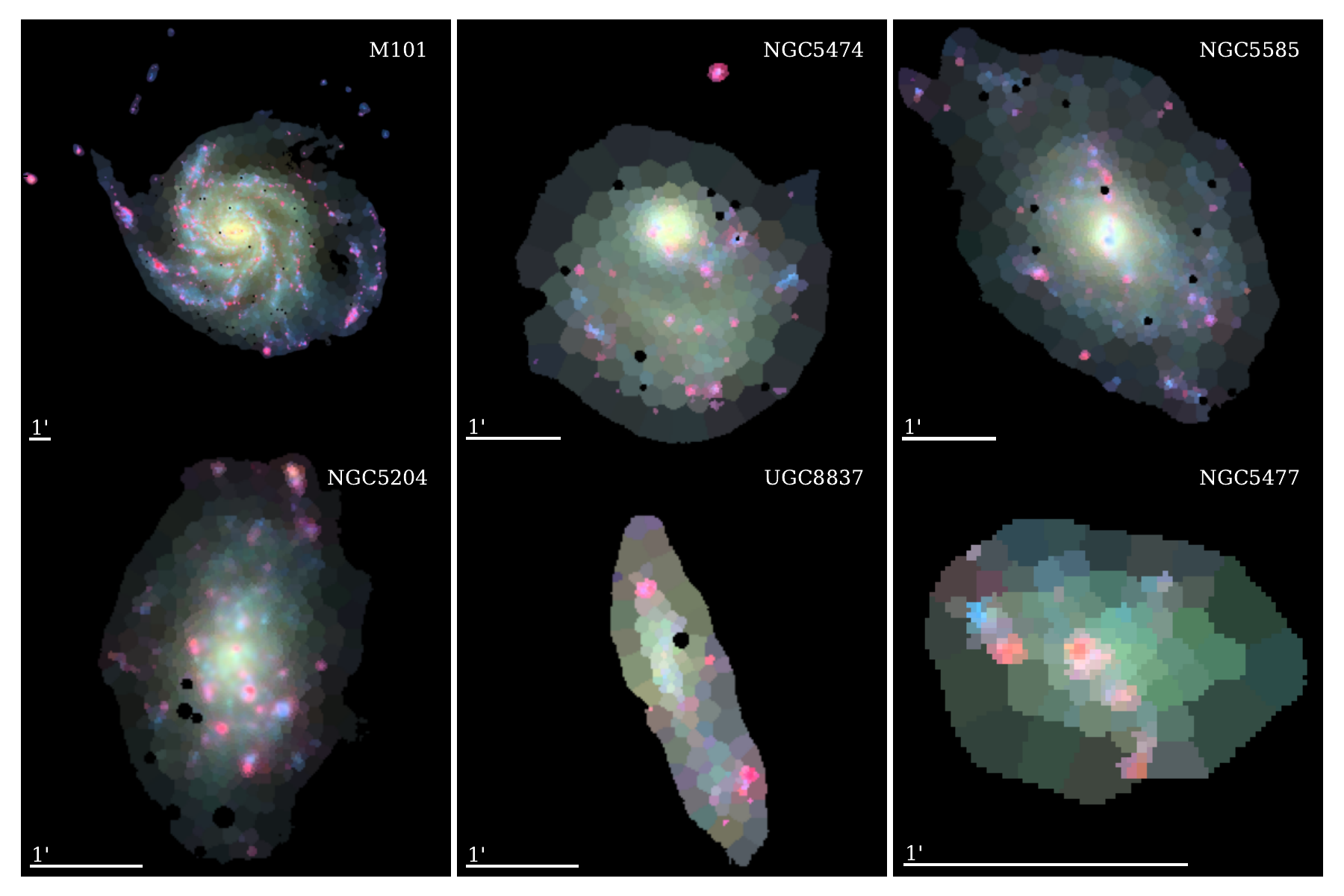}
    \caption{The same that Fig. \ref{fig:groupM101_original} but for the data after the pre-processing.}
    \label{fig:groupM101_posprocessing}
\end{figure*}

Fig.\ \ref{fig:pre-processing_steps_M101} uses the \textit{u} and \textit{r} band images of M101 to illustrate the sequence of pre-processing operations described in Sec.\ \ref{sec:sample}.  
It is notable in Fig. \ref{fig:pre-processing_steps_M101} that the original data in the \textit{u}-band are quite faint, exhibiting a significant number of negative flux pixels (in white in the image), approximately 36\% of those within the galaxy mask, which implies an even higher count of pixels with bad signal. In contrast, the \textit{r}-band, which has a superior S/N ratio, shows 6\% of negative flux pixels. The figure demonstrates the sequential reduction in negative flux pixels after each pre-processing step, which can be read as a measure of improvement in images. Fig.\ \ref{fig:groupM101_posprocessing} shows how the composites in  Fig.\ \ref{fig:groupM101_original} change after pre-processing.

\subsection{Photo-spectral fits and $\overline{\Delta}$ maps}

\begin{figure*}
    \centering
    \includegraphics[width=\linewidth]{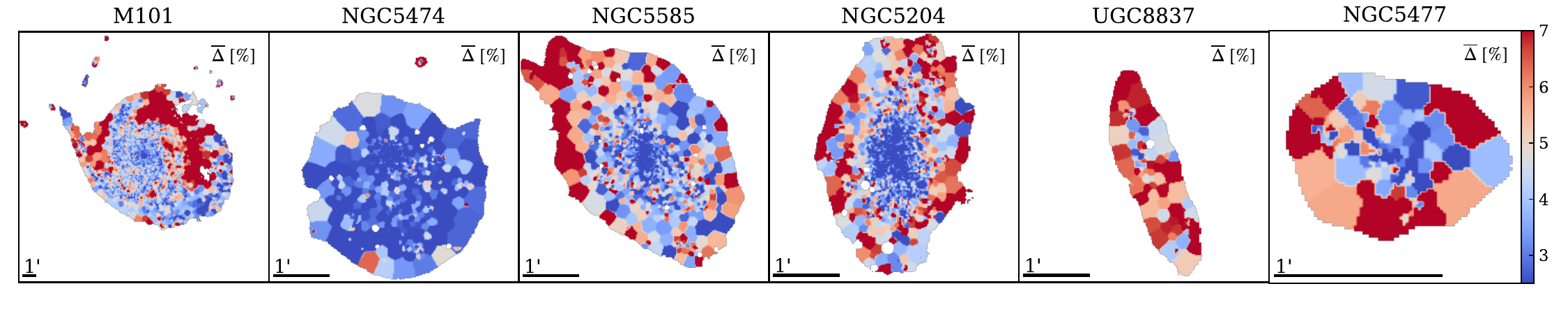}
    \includegraphics[width=0.49\linewidth]{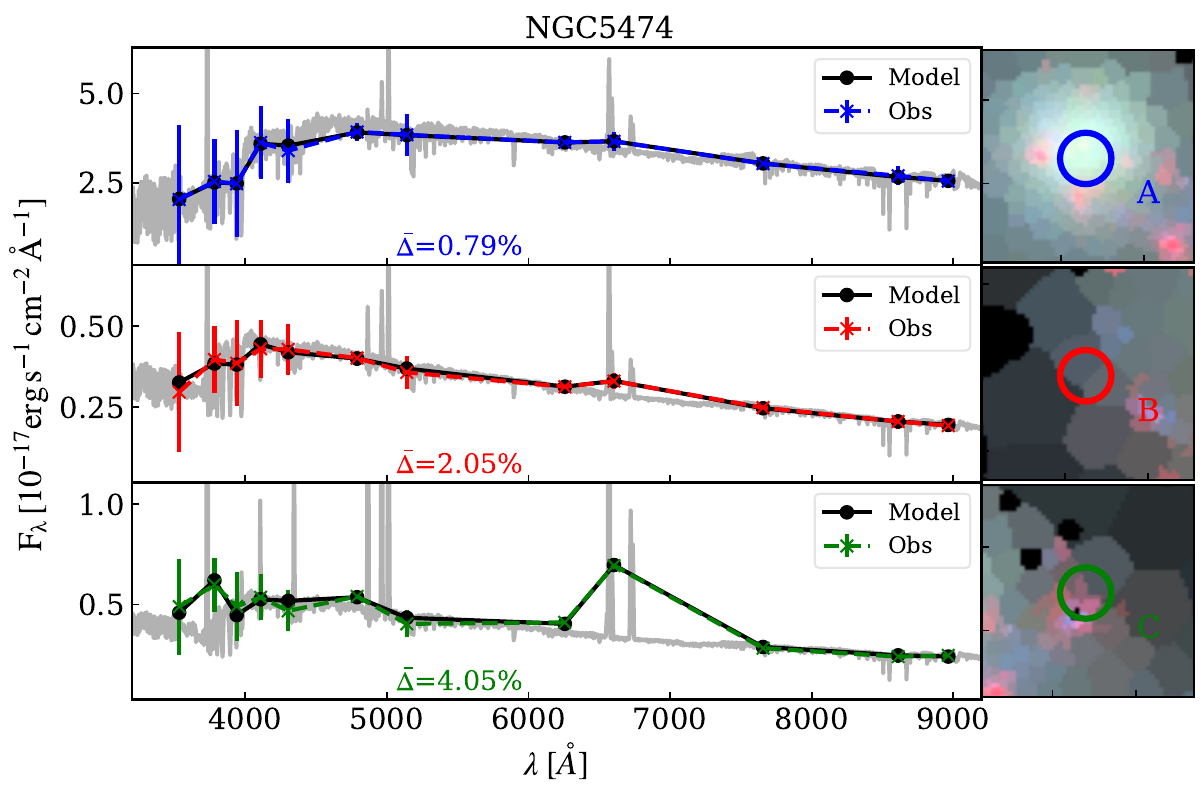}
    \includegraphics[width=0.49\linewidth]{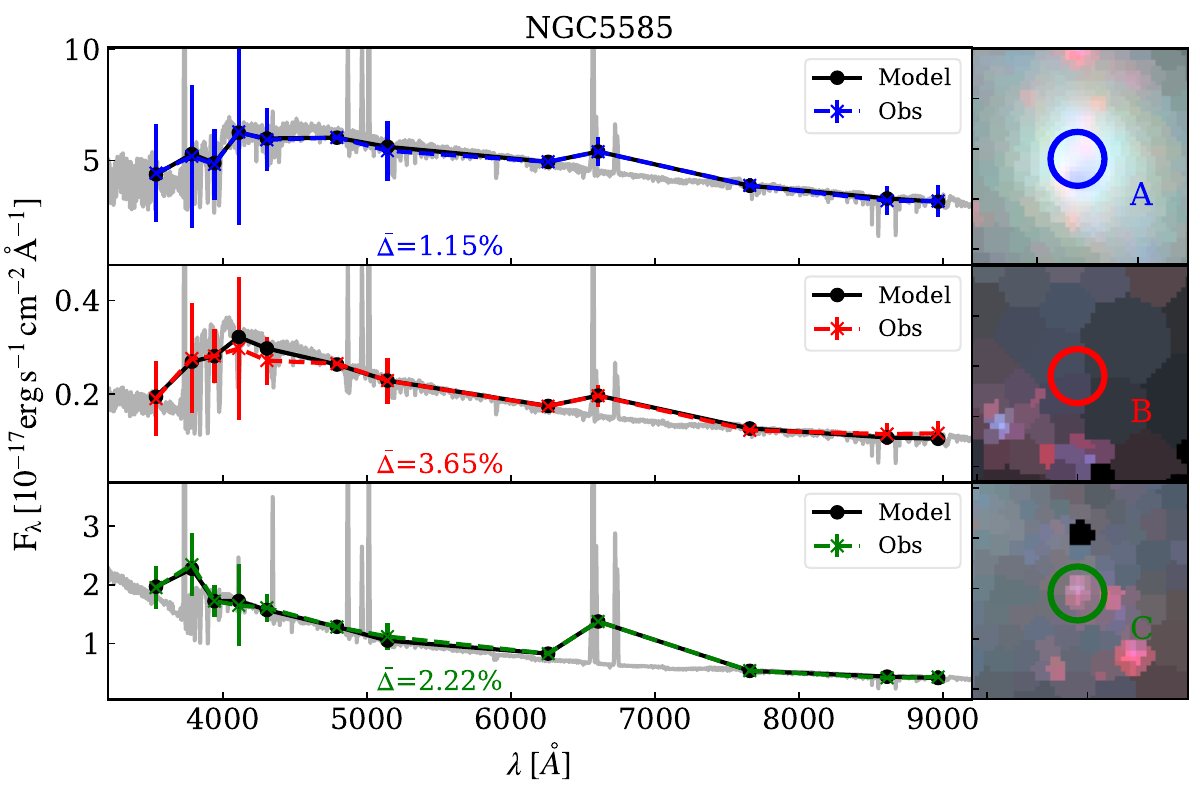}
    \includegraphics[width=0.49\linewidth]{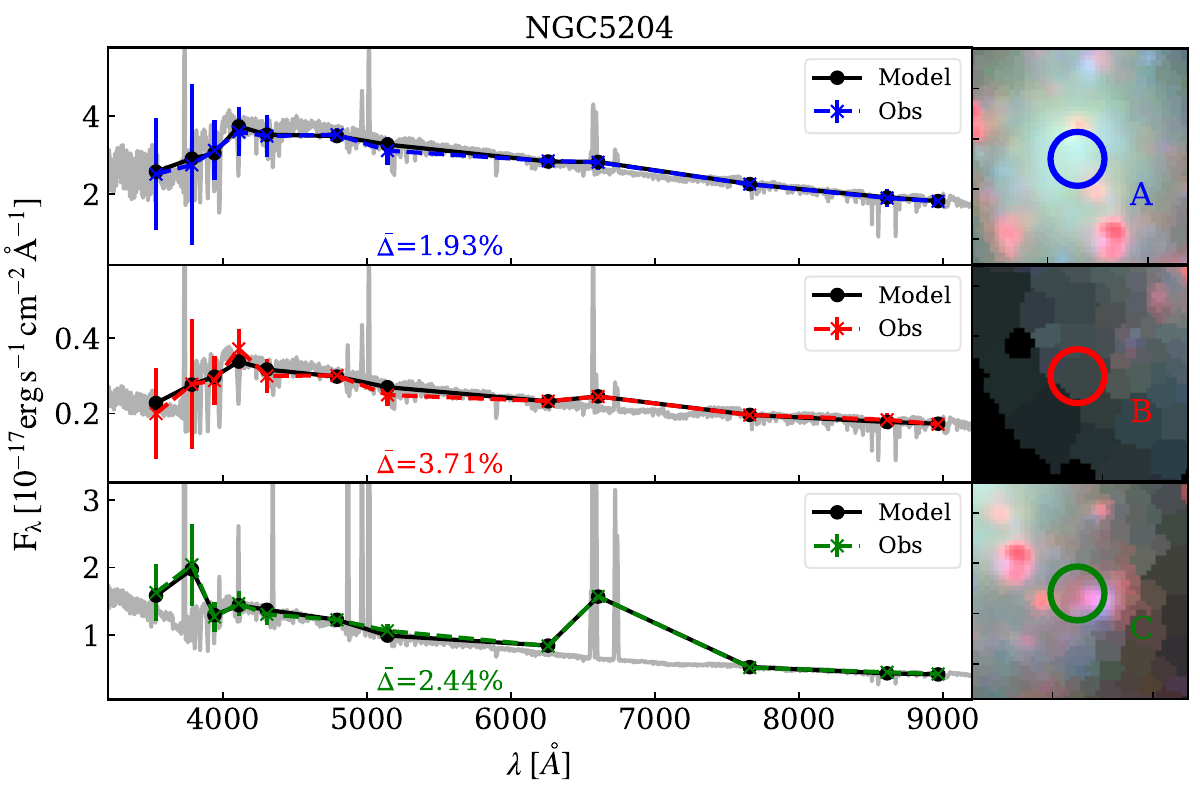}
    \includegraphics[width=0.49\linewidth]{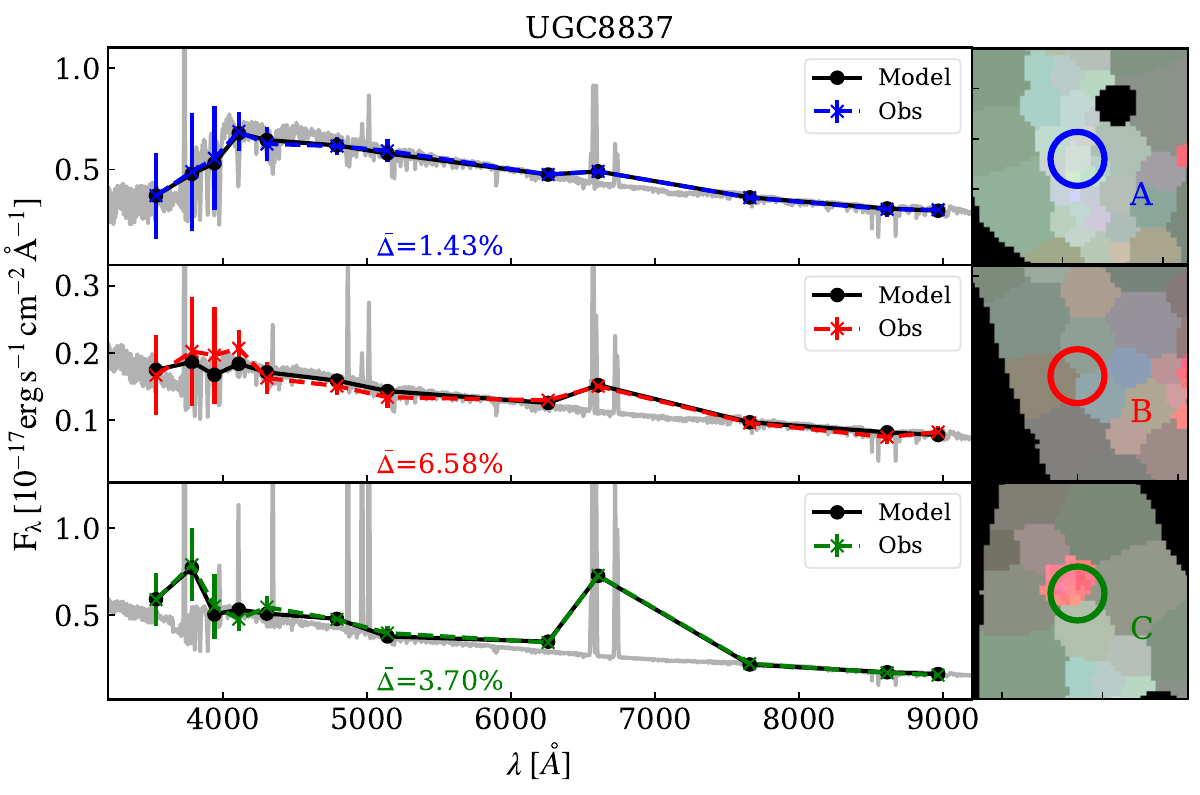}
    \includegraphics[width=0.49\linewidth]{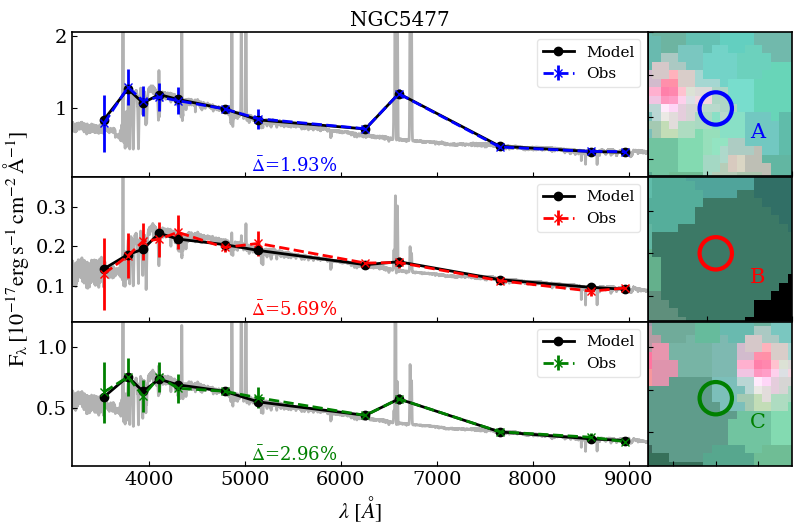}
    \includegraphics[width=0.49\linewidth]{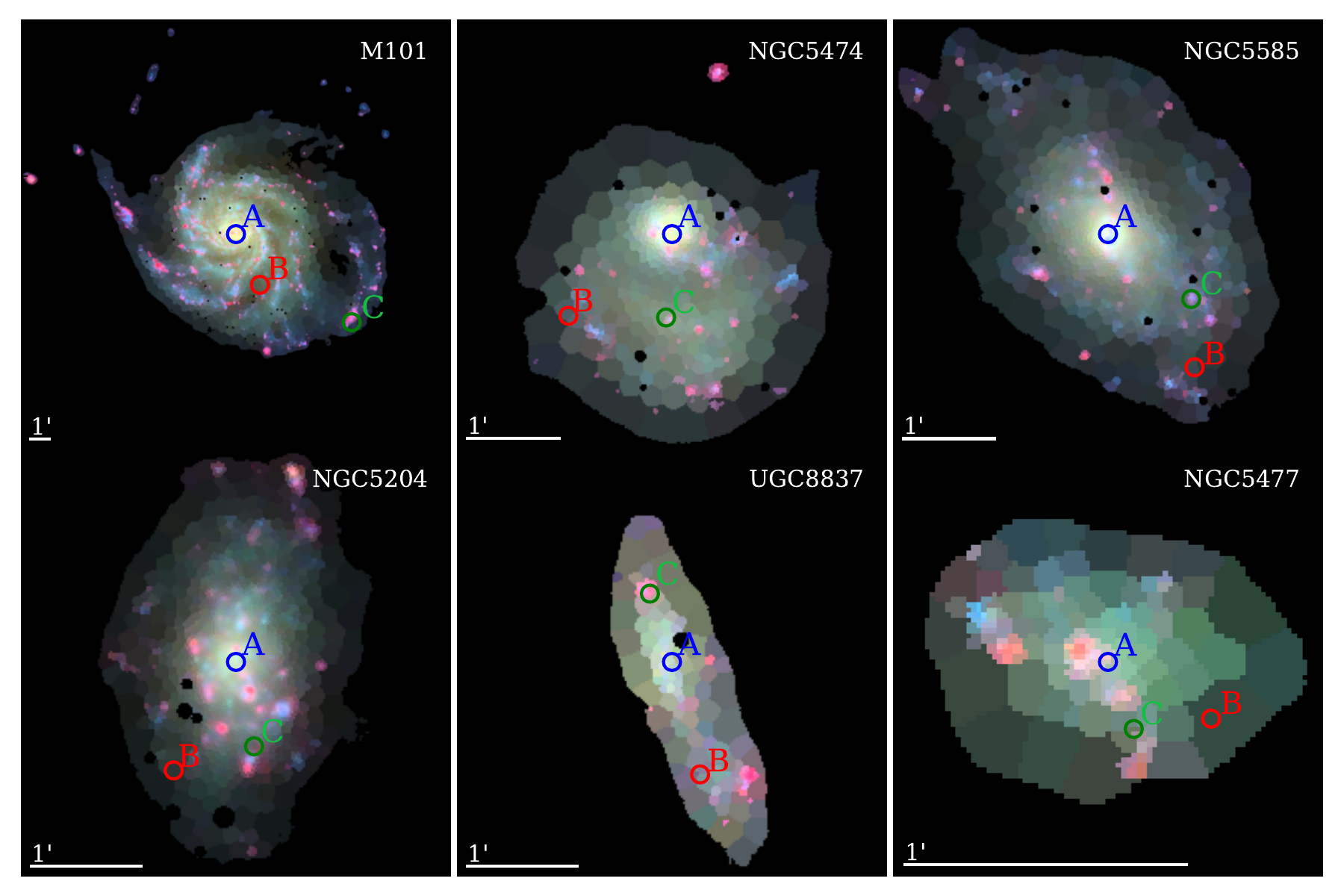} 
    \caption{Maps of the mean relative absolute deviation between observed and model fluxes, $\overline{\Delta} \equiv \langle |O_\lambda - M_\lambda| / M_\lambda \rangle$. Fits are presented as Fig.\ref{fig:fit_M101example}, but for the other galaxies. The last panel shows the fit positions (A, B, and C) in the galaxy.}
    \label{fig:fits_adev_extrasM101group}
\end{figure*}

Like Fig.\ \ref{fig:fit_M101example} in the main text, Fig.\ \ref{fig:fits_adev_extrasM101group} shows three example fits for zones in galaxies other than M101. Region A in all cases shows the fit for the central spaxel. The resulting $\overline{\Delta}$ range from 0.79 to 1.93\%. Region B represents fainter regions, near the boundary of our galaxy mask (as seen in the last panel). As expected, these B-fits have the worst $\overline{\Delta}$, though it is not too bad for the NGC 5474, NGC 5585, and NGC 5204 (the most massive in the figure), varying from 2.05 to 3.75\%. However, for UGC 8837, the noisiest galaxy in the sample, region B has residuals of  $\overline{\Delta}=6.58$\%, a good example of a bad fit. Lastly, the regions labeled C show examples of \hii\ regions, which show significant \Ha\ and \nii\ fluxes in the \textit{J0660} filter, and, in some cases, a prominent \oii\ contribution to the \textit{J0378} filter.

Fig.\ \ref{fig:fits_adev_extrasM101group} shows maps of $\overline{\Delta}$, the mean relative absolute deviation between data and model fluxes, for all galaxies in the sample. Red colors in these maps indicate relatively bad fits ($\overline{\Delta} > 6\%$). Properties derived from these zones should be treated with care. It is notable that, despite the generally good fits of M101, the inter-arm areas exhibit the poorest $\overline{\Delta}$, which is expected given that these areas have less signal. NGC 5474 is the galaxy with the smallest $\overline{\Delta}$ values, with few regions where $\overline{\Delta}\geq6$\%. Both NGC 5585 and NGC 5204 demonstrate good fits, with $\overline{\Delta} \sim 2\%$ at the center, increasing towards the edges as the signal wanes. The dwarfs UGC 8837 and NGC 5477 exhibit a similar pattern.

\subsection{Dust maps and dereddened images}

\begin{figure*}
    \centering
    \includegraphics[width=0.75\linewidth]{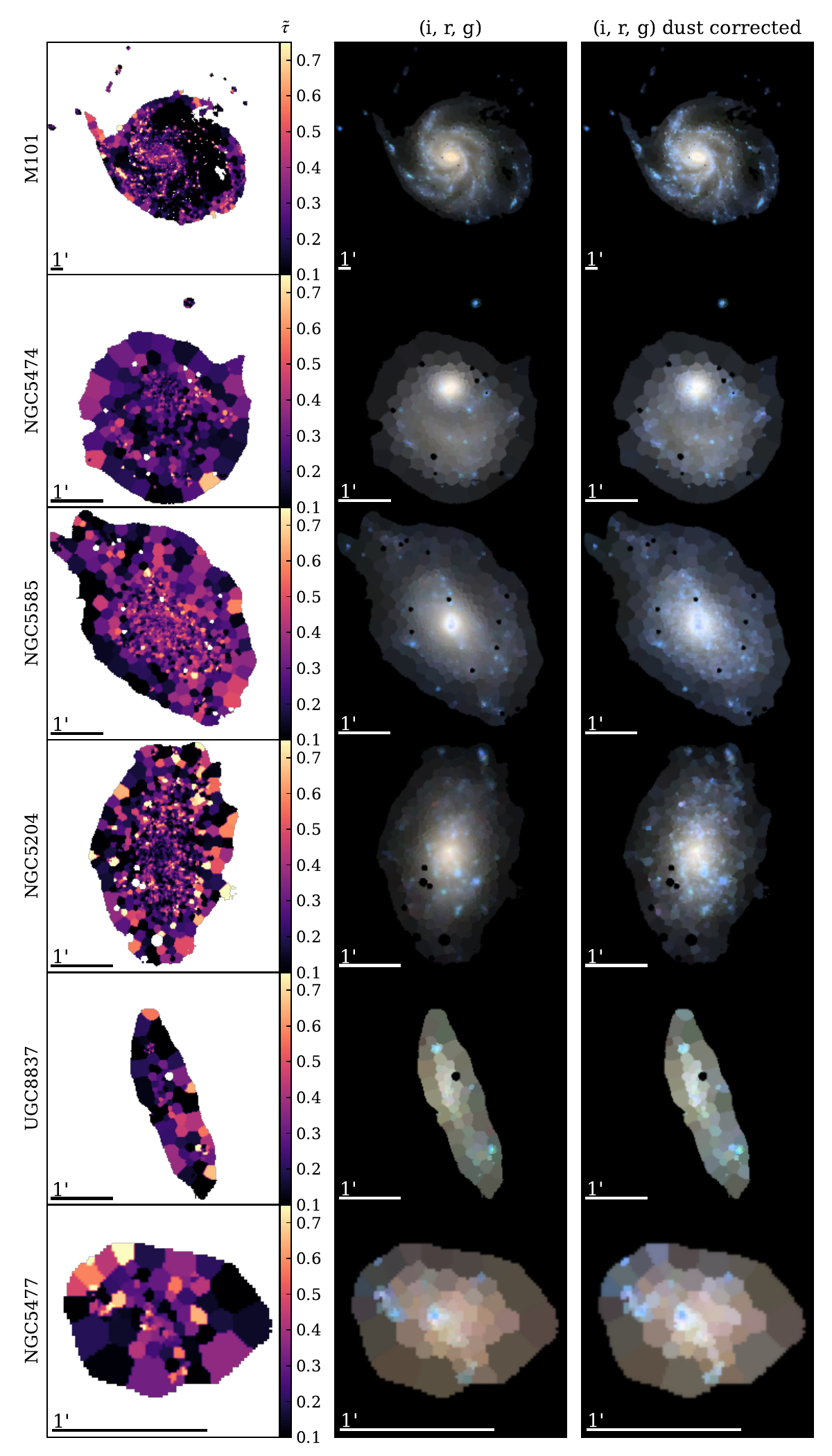}
    \caption{Left: Maps of the effective V-band dust optical depth ($\tilde{\tau}$). The central and right panels show RGB composites with the \textit{i, r}, and \textit{g} bands before and after correction for dust, respectively.
    }
    \label{fig:dust_all_gals_M101group}
\end{figure*}

The left panels in Fig.\ \ref{fig:dust_all_gals_M101group} show maps of the effective V-band dust optical depth $\tilde{\tau}$, defined as 

\begin{equation}
\label{eq:tauEff}
\tilde{\tau} \equiv x_{\rm BC} (\tau^{\rm ISM} + \tau^{\rm BC}) + (1 - x_{\rm BC}) \tau^{\rm ISM}
= x_{\rm BC} \tau^{\rm BC} + \tau^{\rm ISM}
\quad,
\end{equation}

\noindent where $x_{\rm BC}$ is the sum of all $x_j$ fractions associated to populations with age $t_j < t_{\rm BC} = 10$ Myr .
In our two-$\tau$'s model, these young stars are attenuated by $\tau^{\rm ISM} + \tau^{\rm BC}$, while older ones are attenuated by just $\tau^{\rm ISM}$, so that $\tilde{\tau}$ is a convenient weighted average which condenses these two dust components onto a single number. By construction, $\tilde{\tau} \rightarrow \tau^{\rm ISM}$ when young populations are not present, and to $\tau^{\rm ISM} + \tau^{\rm BC}$ when they dominate.

In M101 our dust map follows the spiral arms, with peaks reaching $\tilde{\tau} \sim 1$ aligning with its star-forming regions and a slight general decrease towards large radii. An inner more diffuse component is also seen in the central parts, as well as a horizontal inner bar-like feature like that seen in CO \citep{1991Kenney}. Except for these regions, overall our fits indicate that there is relatively little dust attenuation in M101, consistent with its $\sim$ face-on orientation.
The equivalent $A_V$ values compare relatively well with those derived by \cite{Lin_2013} and \cite{2017_Watkins_Mihos} with very different methodologies.

Other galaxies in the group also show relatively little dust. The $\tilde{\tau}$ maps are patchy, with no clear structure, with peaks in the star forming regions (compare with the $\langle \log t \rangle_L$ and $W_{\Ha}$ maps). Note that we find very little dust in UGC 8837, despite its edge-on orientation.

TB23 used their \alstar\ fits to correct S-PLUS images of NGC 1365 for dust. Our galaxies are not as dusty as NGC 1365, whose inner regions are heavily obscured, but it is still worth repeating this exercise for our J-PLUS data on galaxies in the M101 group. Fig.\ \ref{fig:dust_all_gals_M101group} shows the results. Its middle panels show composites made with the i, r, and g images in the  R, G, and B channels, respectively, while the panels on the right show the same composite after correcting for dust attenuation. 

For M101, the dusty filaments along the inner arms, visible as red stripes in the original image, are absent in the corrected one, indicating that, despite its simplicity, our modeling is able to account for dust attenuation adequately. Moreover, the star-forming regions, where the largest values of $\tilde{\tau}$ are found, now appear bluer, brighter, and with a stronger contrast to the underlying disk. For the other galaxies, where the dust is less structured, the middle and right panels differ in their overall blueness and in the brightness of the star forming regions.

\subsection{Emission lines}

\begin{figure*}
    \centering
    \includegraphics[width=0.89\linewidth]{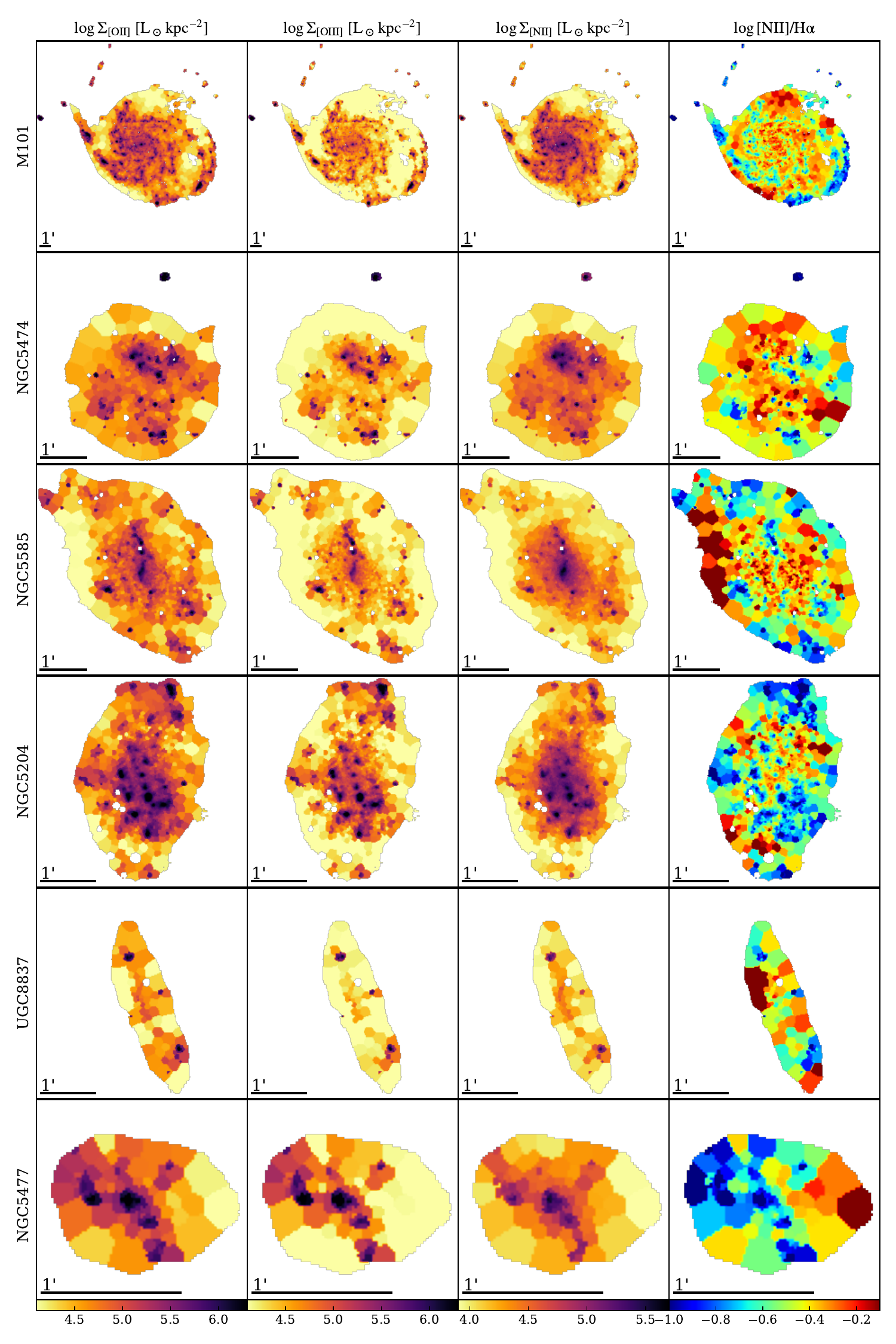}
    \caption{From left to right: Maps of the \oii3727, \oiii5007, and \nii6584 surface brightness, and the \nii/\Ha\ ratio.
    }
    \label{fig:ELs_extrasM101group}
\end{figure*}

Fig.\ \ref{fig:ELs_extrasM101group} shows complementary maps related to our EL results. 
The $\Sigma_{\Ha}$ maps were previously shown in Fig.\ \ref{fig:ELPropMaps_M101Group}, while Fig.\ \ref{fig:ELs_extrasM101group} shows the surface brightness maps for the forbidden lines \oii, \oiii\ and \nii, as well as the \nii/\Ha\ flux ratio.
Note that, for the redshifts of galaxies in the M101 group, only \oii\ and \nii\ + \Ha\ fall under narrow bands (\textit{J0378} and \textit{J0660}). Yet, as demonstrated in TB23, because of the empirical nature of the EL base used in our fits, even lines covered only by broad bands are well recovered by our method.

As in the \Ha\ maps, the \oiii\ maps in Fig.\ \ref{fig:ELs_extrasM101group} also show the star-forming regions standing out clearly and in sharp contrast with their surroundings. These same regions stand out in the \oii\ and \nii\ maps, although these are visibly more diffuse than the \oiii\ and \Ha\ ones. These low ionization lines are characteristic of DIG emission.

The \nii/\Ha\ map for M101 shows this EL flux ratio decreasing from $\sim 0.5$ in the inner star-forming regions to less than 0.1 in the outer ones. As suspected from Fig.\ \ref{fig:ELs_extrasM101group}, in the central parts of M101, as well as in between its spiral arms, \nii/\Ha\ is larger, reaching values typical of DIG. As discussed in Sec. \ref{sec:ELPropMaps_M101Group}, the \WHa\ map in the second column panel of Fig.\ \ref{fig:ELPropMaps_M101Group} confirms this interpretation.
A gradient in \nii/\Ha\ is also present in NGC 5474, NGC 5585 and NGC 5204, though not as strong as in M101, while no clear trend is identifiable in the two dwarf galaxies.

\subsection{Uncertainties}
\label{sec:Uncertainties}

This final section addresses the uncertainties in the main properties derived from our methodology. 
As one can infer from the overall consistency of the results, as well as from the reported agreement with previous studies, uncertainties are not a limiting factor in our analysis. It is, nonetheless, relevant to discuss them, specially given some of the novelties in our approach (like the modeling of ELs). Though based on the results for galaxies in the M101 group, the discussion below serves as a guide for future studies where the same kind of analysis is applied to other targets.

\subsubsection{Uncertainties in individual fits}

The dispersion among the MC runs in \alstar\ offers a way to estimate the uncertainties in the derived properties. The experiments in TB23, where these dispersions were compared to the dispersion of the difference ($\delta$) between properties derived from \starlight\ full spectral fits of SDSS galaxies and those obtained from \alstar\  fits of the corresponding synthetic photometry in the S-PLUS bands, confirm that this is a reasonable method to estimate uncertainties, even though a tendency was noted for the \alstar-based MC dispersions to overestimate the more heuristic $\delta$-based ones by factors of up to 2, depending on the property.
Let us thus examine the MC-based dispersions in the properties derived for individual fits, be they single pixels or Voronoi zones. We will use the $\sigma_{\rm NMAD}$ statistic as a measure of dispersion.

The median values of $\sigma_{\rm NMAD}$\  for $\log \Sigma_\star$ span the 0.22--0.25 dex range for the six galaxies in the M101 group.
In star-forming regions, however, $\sigma_{\rm NMAD}$ can reach values of 0.4 dex or larger. 
These regions have their light completely dominated by young populations, but even a 1\% light fraction due to old stars would be enough to make them dominant in mass. Variations in this very poorly constrained old component among the MC fits explain the large spread in $\Sigma_\star$ in these regions. As an illustration, the median $\sigma_{\rm NMAD}(\log \Sigma_\star)$ in M101 changes from 0.17 dex for regions where $W_{\Ha} < 30$ \AA\ to 0.26 dex for those where \Ha\ has a larger equivalent width.

\begin{figure*}
    \centering
    \includegraphics[width=\linewidth]{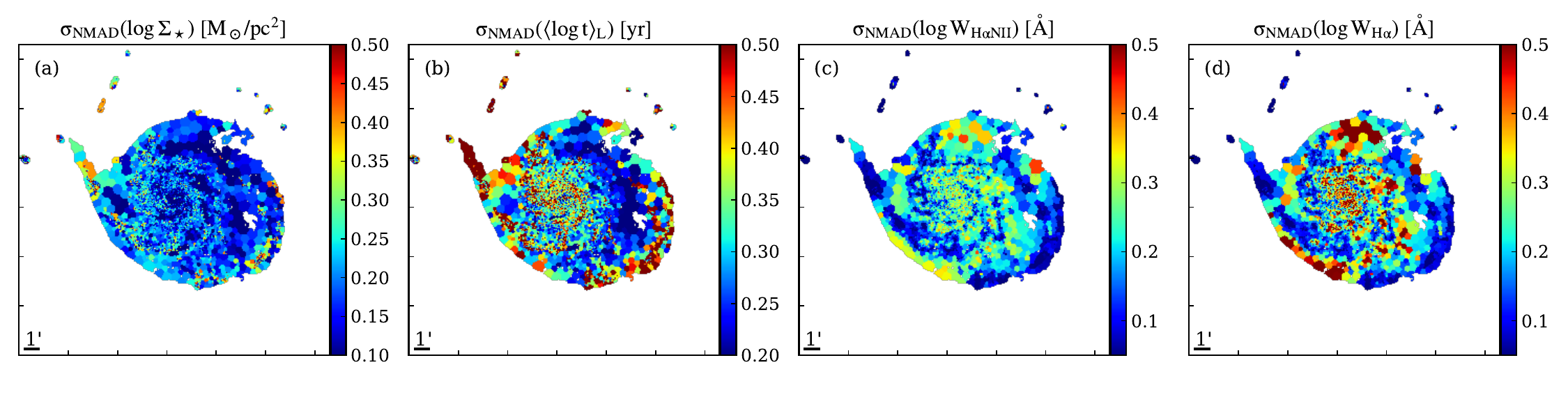}
        \caption{Maps of the $\sigma_{\rm NMAD}$ dispersion among the Monte Carlo runs for $\log \Sigma_\star$, $\langle \log t \rangle_L$,  $\log W_{\Ha \mathrm{N}\,\textsc{ii}}$, and $\log W_{\Ha}$ for M101.}
    \label{fig:Uncertainties_M101}
\end{figure*}

Fig.\ \ref{fig:Uncertainties_M101}a shows the map of $\sigma_{\rm NMAD}(\log \Sigma_\star)$ for M101. Comparing this map with the ones for $\langle \log t \rangle_L$ or $W_{\Ha}$ (Figs.\ \ref{fig:StellarPropMaps_M101Group} and \ref{fig:ELPropMaps_M101Group}) one sees low $\sigma_{\rm NMAD}$ in regions dominated by old populations, and larger uncertainties in regions of ongoing star formation, as summarized above.

Regarding mean ages, we find median values of $\sigma_{\rm NMAD}(\langle \log t \rangle_L)$ in the 0.34--0.52 dex range for our galaxies.  $\sigma_{\rm NMAD}(\langle \log t \rangle_L)$ tends to be larger for regions with ages close to the middle of the age grid (at $\sim 10^8$ yr), where the larger number of combinations of different components leading to approximately the same total spectrum inevitably translates to a larger dispersion in $\langle \log t \rangle_L$. The map of $\sigma_{\rm NMAD}(\langle \log t \rangle_L)$ in Fig.\ \ref{fig:Uncertainties_M101}b shows a general progression towards larger uncertainties towards the outer regions. The red regions, where $\sigma_{\rm NMAD} \ge 0.5$ dex, are mostly located in regions of intermediate age and $W_{\Ha}$, often around \hii\ regions (see Figs.\ \ref{fig:StellarPropMaps_M101Group} and \ref{fig:ELPropMaps_M101Group}). In the \hii\ regions themselves uncertainties are smaller, though not as small as in zones dominated by older populations. Overall, though acceptable for photometric work, our uncertainties in 
$\langle \log t \rangle_L$ are substantially larger than those attainable with spectroscopy, which are typically about 0.1 dex (e.g., \citealt{CidFernandes2014}). 

Uncertainties in dust attenuation are also larger than the typical $\pm 0.1$ mag in $A_V$ attainable with full spectral synthesis over a similar spectral range. We obtain median $\sigma_{\rm NMAD}(\tau^{\rm ISM})$ values of 0.15--0.22 for our galaxies, corresponding to 0.16--0.24 mag in $A_V$. Stellar metallicities, on the other hand, turn out to have small uncertainties ($\sim 0.2$ dex), but this is mainly due to the restricted range in $Z$ allowed for in the fits. 
Regarding SFRs, we obtain median $\sigma_{\rm NMAD}(\log \Sigma^\star_{\rm SFR})$ values of 0.15--0.22 dex across our sample. These dispersions are obtained after discarding zones where $< 100$ Myr stars contribute less than 30\% to the flux at 5635 \AA, i.e., by focusing on regions where these populations are reliably detected.

Turning to EL properties, let us first examine the combined equivalent widths of the \Ha\ and $\nii\lambda\lambda6548, 6584$ lines, $W_{\Ha \mathrm{N}\,\textsc{ii}}$.  
We find median $\sigma_{\rm NMAD}(\log W_{\Ha \mathrm{N}\,\textsc{ii}})$ values of 0.05--0.25 dex over our sample (see Fig.\ \ref{fig:Uncertainties_M101}c). The highest value of 0.25 dex is for NGC 5585, due to its unusual noisy data in the \textit{J0660} filter, with photometric uncertainties about $\sim 3.3\times$ larger than in the other galaxies.

Considering only zones where $W_{\Ha \mathrm{N}\,\textsc{ii}}>30 \ \AA$, the median values of $\sigma_{\rm NMAD}(\log W_{\Ha \mathrm{N}\,\textsc{ii}})$ for all galaxies is just $\sim 0.05$ dex. Such small uncertainties are not surprising, given that, at the redshift of our sources, these lines all fall within the \textit{J0660} filter, so that their combined equivalent width can be reliably estimated. This is in fact the reason why in TB23 we have devised a whole empirical $W_{\Ha \mathrm{N}\,\textsc{ii}}$-based scheme to disentangle \Ha\ from \nii\, which greatly improves the EL diagnostic power of S-PLUS and J-PLUS data.

In itself, however, $W_{\Ha \mathrm{N}\,\textsc{ii}}$ is not a particularly interesting index.
$W_{\Ha}$, on the other hand, is a very useful tracer of specific SFR (or age of an \hii\ region, depending on the scale one is looking at). We obtain median $\sigma_{\rm NMAD}(\log W_{\Ha})$ values of $\sim 0.1$ dex for M101, NGC 5204, UGC 8837, and NGC 5477, while for NGC 5474 the value is 0.2 dex and for NGC 5585 it is 0.3 dex. In all cases, the relative uncertainty (i.e., $\sigma_W/W$) decreases as $W_{\Ha}$ increases. Even when $W_{\Ha}$ is small (and thus uncertain) we are able to identify is as small, which is good enough for basic diagnostics, like distinguishing DIG from \hii\ regions.

Fig.\ \ref{fig:Uncertainties_M101}d shows the map of $\sigma_{\rm NMAD}(\log W_{\Ha})$ for M101. Comparing to 
Fig.\ \ref{fig:ELPropMaps_M101Group}, one sees that $\log W_{\Ha}$ is most uncertain in regions where
$W_{\Ha}$ is just a few \AA. In fact, up to $W_{\Ha} \sim 10$ \AA\  the MC dispersion is of the order of $W_{\Ha}$ itself. Conversely, uncertainties in $\log W_{\Ha}$ are very low in star-forming regions.

Another EL property of interest is the \nii/\Ha\ ratio, which, besides being a nebular metallicity indicator by itself, has a major diagnostic role in separating star-forming regions from those where other sources of ionization are present. The median $\sigma_{\rm NMAD}(\log \nii/\Ha)$ values for our galaxies are all very close to 0.2 dex, and in very few regions it reaches 0.3 dex. This apparently good precision (for photometric data) is, however, in most part built into our method, which uses an initial estimate of $W_{\Ha \mathrm{N}\,\textsc{ii}}$ to restrict the possible EL-space available to the fits.

Finally, the MC uncertainties on $12 + \log {\rm O/H}$ run from 0.05 to 0.2 dex over the body of our galaxies, increasing systematically as O/H increases. This happens because our empirical EL-base is built to span the observed BPT diagram, which is narrower at its top-left, low O/H regime, than at its center-bottom, where O/H is larger. Consequently, the MC fits mix fewer base elements at low O/H than at high O/H. 
Notice that these uncertainties are of the same order of the overall variations in $\log O/H$ within our galaxies. This, however, does not challenge the significance of the nebular MZRs seen in Fig.\ \ref {fig:MedianRelations4M101group}, since they were obtained not on the basis of individual fits, but from median curves over many such fits, and thus much less uncertain them each of them. 
Statistics, in fact, plays a major role in the assessment of uncertainties, as discussed next.

\subsubsection{Uncertainties in practice} 

The uncertainties in the individual fits to J-PLUS photometry discussed just above are clearly larger than those ones grew accustomed to in the era of spectroscopic surveys, as expected from the disparity in information content between, say, the thousands of flux entries in an SDSS or CALIFA spectrum and the 12 points ``summary of a spectrum'' offered by J-PLUS. 

In practice, however, both in bona fide IFS and IFS-like studies, one never focuses on results for an individual spaxel (or Voronoi zone). Instead, spatial averages are always considered in one way or another. This averaging can be either explicit, like when one examines radial profiles (including alternative representations like those in Fig.\ \ref{fig:age-Z-Mass_M101group}, where the $x$-axis is only indirectly related to radius), or implicit, like when one inspects maps like those in Figs.\ \ref{fig:StellarPropMaps_M101Group} and \ref{fig:ELPropMaps_M101Group}. This averaging process reduces the formal uncertainties of individual fits by factors of $\sqrt N$, which in most circumstances make them negligible.

There are, of course, other less formal sources of uncertainty inherent in this kind of analysis. For instance, we adopted a fixed dust attenuation law and a pre-defined recipe for differential extinction, which is at best an approximation to reality. Our choice of spectral models for stellar populations, with its built-in assumptions and approximations, is also a source of uncertainty, as is the way we define our EL base as well as our implementation of an empirical EL prior to disentangle \nii\ from \Ha. In summary, each of these modeling choices carries its own set of assumptions, contributing to the overall uncertainty in the analysis.

While we have not quantified these modeling-related uncertainties in this work, we can borrow from the thorough analysis of uncertainties in full spectral fitting of CALIFA datacubes by \cite{CidFernandes2014}. They find that different choices of stellar population models impact the results more than uncertainties related to the method itself, a conclusion that most likely applies to this work too. Substantial changes in the results reported here are thus more likely to come from changes in the ingredients of the analysis or eventual recalibrations of the data.

\end{appendix}
\end{document}